\documentclass[aps,pre,twocolumn,showpacs,floatfix]{revtex4}
\usepackage{amsmath,bm,epsfig}
\usepackage{latexsym}
\usepackage{amssymb}
\usepackage{color}
\usepackage{morefloats}
\usepackage{subfigure}
\usepackage{graphicx}
\usepackage{amsmath}
\usepackage{bm}
\usepackage{float}
\usepackage{makeidx}
\definecolor{revised}{RGB}{0,0,0}
\definecolor{new}{RGB}{0,0,0}

\begin{document}

\title{Emergence of a dynamical state of coherent bursting with power-law distributed avalanches from collective stochastic dynamics of adaptive neurons}
\author{Lik-Chun Chan, Tsz-Fung Kok, and Emily S.C. Ching\footnote{Corresponding author; email address: ching@phy.cuhk.edu.hk}}
\affiliation{Department of Physics, The Chinese University of
Hong Kong, Shatin, Hong Kong}
\date{\today}
\begin{abstract}
Spontaneous brain activity in the absence of external stimuli is not random but contains complex dynamical structures such as
neuronal avalanches with power-law duration and size distributions. These experimental observations have been 
interpreted as supporting evidence for the hypothesis that the brain is operating {\color{revised} near a critical point of phase transition between two states,} and attracted much attention. 
Here, we show that a dynamical state of coherent bursting, with power-law distributed avalanches and features as observed in experiments, emerges in networks of adaptive neurons with stochastic input 
when excitation is sufficiently strong and balanced by adaptation. We demonstrate that these power-law distributed avalanches are direct consequences of 
stochasticity and coherent bursting, which in turn is the result of a balance between excitation and adaptation. 
{\color{revised} Our work thus shows that the observed dynamical features of neuronal avalanches can arise from collective stochastic dynamics of adaptive neurons under suitable conditions and need not be signatures of criticality.}
\end{abstract}  \maketitle

\section{Introduction}\label{intro}

Neurons in the brain fire even when they do not receive any external stimuli and such firing activities are known as spontaneous activity (see reviews~\cite{RM2006,Buzsaki2019} and references therein).
Spontaneous activity is far from random but has rich dynamical structures in space and time. There are low frequency oscillations in the range of 0.01 - 0.2~Hz~\cite{HS1997,Krishnan2018}.  
Collective bursting events at time scales shorter than about 100~ms, known as neuronal avalanches, have been revealed from
 analyses of in vitro measurements from cortical slices and dissociated cultures of cortical 
neurons of rats~\cite{BeggsPlenz2003,pone2007,Pasquale2008} and
in vivo measurements from anesthetized rats~\cite{Gireesh2008} and awake monkeys~\cite{Petermann2009}. 
These neuronal avalanches have power-law distributions in duration $T$ and size $S$:
\begin{eqnarray}
P(T) &\sim& T^{-\tau_T} \label{duration}\\
P(S) &\sim& S^{-\tau_S}  \label{size}
\end{eqnarray}
{\color{new} where $P(X)$ denotes the probability distribution of $X$ and $\sim$ indicates proportionality.}
It has further been found~\cite{ Friedman2012} that the mean temporal profiles of avalanches of varying durations can be described by a scaling function 
such that the average size of the avalanches {\color{new} conditioned on} a given duration $T$ has a power-law dependence on $T$:
\begin{equation}
\langle S | T \rangle \sim T^\gamma
\label{gamma}
\end{equation}
{\color{new} where $\langle X | Y \rangle$ denotes the conditional average of $X$ given the value of $Y$. Moreover,}
the exponents $\tau_T$, $\tau_S$ and $\gamma$  obey approximately a scaling relation 
\begin{equation} 
 \frac{\tau_T -1 }{\tau_S -1}  =  \gamma 
 \label{relation}
\end{equation}
This scaling relation has been derived for the phenomenon of crackling noise~\cite{crackling2001,Sethna2006}.

{\color{new} The observed power-law distributed neuronal avalanches, with an approximate scaling relation between the exponents, in spontaneous activity have attracted particular attention as they have been interpreted as evidence supporting the hypothesis that the brain is operating near a critical point between two states~~\cite{Beggs2008,Chialvo2010,BeggsTimme2012,Cocchi2017,BJ2022}.}
{\color{revised} From the study of continuous phase transitions of physical systems~(see, e.g.,~\cite{Stanley1971,Goldenfeld1992,Nishimori2011}), 
it is known that when a system approaches a critical point of a phase transition between two states, various physical variables diverge or go to zero in the form of power laws.
These power-law exponents are related by certain scaling relations and are characteristics of the system. Power-law distributions are also commonly found in nature, for example, in earthquakes, snow avalanches and crackling noise. 
In their seminal papers~\cite{SOC1987,SOC1988}, Bak, Tang and Wiesenfeld introduced the concept of self-organized criticality to show how extended systems of many degrees of freedom can naturally reach a critical point and generate avalanches {\color{new} that have power-law size and duration distributions.}} {\color{revised} The critical brain hypothesis suggests that operating near a critical point can bring functional benefits
such as optimized performance in information processing and transmission~\cite{KC2006,Shew2009,Shew2011,Boedecker2012,ShewPlenz2013,Shriki2016}. While this appealing idea  has inspired extensive studies~\cite{PRL2005,NatPhy2007,PRL2009,NatPhy2010,Rubinov2011,PRL2012,PRL2017,PRL2020,PRE2021,PRE2022,Ehsani2023,Wang2011,Poil2012,PNAS2018,PC2019,PRR2021,NatCompSci2023}, it remains controversial.

{\color{new} In the definition of neuronal avalanches, a time scale is involved, which is} the width $\Delta t$ of the time bins used in the analyses of the measurements. The power-law exponents $\tau_T$ and $\tau_S$ were found to depend on $\Delta t$~\cite{BeggsPlenz2003,Petermann2009}, {\color{new} in contrast to the general understanding that power-law exponents at critical point are characteristics of the system.} This issue is often bypassed by {\color{new} reporting results for $\Delta t$ taken} as the average inter-event interval IEI$_{\rm ave}$ of the whole neuronal population~\cite{BeggsPlenz2003}. Even with this choice of $\Delta t={\rm IEI}_{\rm ave}$, 
 dependence of $\tau_T$, $\tau_S$ and $\gamma$ on the spiking variability level of in vivo measurements and on different animal species has further been found~\cite{Fontenele2019}.
Despite these variations, it has been shown that at some spiking variability level, Eq.~(\ref{relation}) holds with 
 $(\tau_T-1)/(\tau_S-1) \approx 1.28$ for different animal species~\cite{Fontenele2019}.
The observed variation of the exponents with spiking variability level has been addressed by a {\color{new} refined} hypothesis that a healthy brain tend to operate near a line of maximal dynamical susceptibility, which departs from a critical point when there 
is external drive such as external stimulus~\cite{Fosque2021,Front2022}. {\color{new} In these studies~\cite{Fosque2021,Front2022},} variations of $\tau_T$ and $\tau_S$ with $(\tau_T-1)/(\tau_S-1)$ ranging from 1.37 to 1.66 could be obtained
using a cortical branching model for quasicritical brain dynamics~\cite{Quasibrain}. This model lacks biological details and contains only excitation, and the values of 
$(\tau_T-1)/(\tau_S-1)$ obtained differ 
from the observed experimental value of 1.28~\cite{Fontenele2019}.

In addition, power-law distributions are not unique to systems near a critical transition point and can be generated by other mechanisms~(see \cite{Newman2005} and references therein). 
Power-law distributed avalanches 
have been shown to exist in stochastic noncritical systems, systems not near a critical point~\cite{TD2010,PLOS2010,FL2013,PRX2017,PRE2017,PS2018,Faqeeh2019}. 
It has also been shown that inhomogeneous Poisson processes with certain time-varying rates can give rise to avalanches of approximate power-law distributions~\cite{PRE2017,PS2018}  {\color{revised} and analytical results for the exponents {\color{new} in the slow rate regime} have been derived~\cite{PRE2017} but they do not satisfy Eq.~(\ref{relation}).
Using the well-studied network of excitatory and inhibitory leaky integrate-and-fire neurons introduced by Brunel~\cite{Brunel2000}, 
Touboul and Destexhe~\cite{PRE2017} showed that neuronal avalanches with size and duration distributions described by power laws with cutoffs and satisfying Eq.~(\ref{gamma}) exist, away from any critical point, in the synchronous irregular state
but the scaling relation Eq.~(\ref{relation}) is not satisfied.}
Hence, the debate on whether or not the brain is operating at criticality is ongoing~\cite{WP2019,DT2021,Beggs2022} 
and the underlying mechanism of the observed oscillations and power-law distributed neuronal avalanches in spontaneous brain activity remain to be fully understood.

{\color{revised} In this paper, we show that a dynamical state of coherent bursting emerges in networks of adaptive neurons subjected to stochastic input. We use the Izhikevich neuron model~\cite{Izhikevich2003}, which captures spike frequency adaptation  of biological neurons that is missing in the leaky integrate-and-fire model. Neuronal avalanches are found to satisfy Eqs.(\ref{duration})-(\ref{relation}) together with $(\tau_T-1)/(\tau_S-1) \approx 1.3$, as observed in
 experiments~\cite{Fontenele2019}, in this entire dynamical state. This state emerges when the excitation is sufficiently strong and yet small enough to be balanced by  adaptation. We show that the power-law distributed avalanches found in this state are direct consequences of stochasticity and the oscillatory firing rate due to coherent bursting. Our work thus shows that the dynamical features of neuronal avalanches observed in spontaneous brain activity can arise from collective stochastic dynamics of adaptive neurons under suitable conditions and need not be signatures of criticality.}

\section{Model}\label{model}
{\color{revised} We study networks of $N$ neurons subjected to stochastic input using the Izhikevich neuron model~\cite{Izhikevich2003} and
 the more realistic} conductance-based synapse model~\cite{Destexhe1998,Meffin2004,Cavallari2014}.
In the Izhikevich model~\cite{Izhikevich2003}, the dynamics of a neuron is described by a membrane potential $v_i$ and a membrane recovery variable $u_i$, where the label $i$ runs from 1 to $N$ in the network.
The recovery variable $u_i$ accounts for the activation of potassium ionic currents and inactivation of sodium ionic currents, and provides a negative feedback to $v_i$. 
Neuron $i$ receives a synaptic current $I^{\rm syn}_i$ from all its pre-synaptic neurons and a stochastic current $I_{\rm noise}$.
The time evolution of $v_i$ and $u_i$ is governed by two coupled nonlinear differential equations
\begin{eqnarray}
\frac{d v_i}{dt} &=& 0.04 v_i^2 + 5 v_i + 140 - u_i + I^{\rm syn}_i + I_{\rm noise} \label{SDE1}\\
\frac{d u_i}{dt} &=& a(b v_i - u_i) \label{SDE2}\end{eqnarray}
where $v_i$ and $u_i$ are in mV, $t$ is time in ms and $a$ and $b$ are positive parameters. {\color{revised} In Eq.~(\ref{SDE1}), the quadratic dependence on $v$ is derived using bifurcation theory and normal
form reduction~\cite{Izhikevich2007} and the specific choice of the coefficients in $0.04v^2+5v+140$ is made to match the spike initiation dynamics of a biological cortical neuron~\cite{Izhikevich2003}.}
The stochastic current {\color{new} is given by}
\begin{equation}
{\color{new} I_{\rm noise}=\alpha \xi , }
\label{Inoise}
\end{equation}
where $\xi$ is a Gaussian white
noise with zero mean and unit variance: 
\begin{eqnarray}
\langle \xi(t) \rangle=0, &\qquad&
\langle \xi(t_1)\xi(t_2) \rangle = \delta(t_1-t_2)
\end{eqnarray}
and $\alpha >0$ measures the strength of stochastic driving.
Every time when $v_i \ge 30$, neuron $i$ fires and a spike is generated, then $v_i$ and $u_i$ 
are reset:
\begin{eqnarray}
\begin{cases}v_i \rightarrow c \\
u_i \rightarrow u_i + d\end{cases}
\label{adapt}
\end{eqnarray}
where $c>0$ and $d>0$ are parameters.
As $d >0$,  $u_i$  and thus
the negative feedback increases every time a neuron spikes and this mimics spike frequency adaptation, 
which is a reduction in the spiking frequency over time in response to a prolonged stimulus, a property observed in biological neurons.
It has been shown that~\cite{Izhikevich2007} with suitable choices of the parameters $a$, $b$, $c$, and $d$, the Izhikevich model 
is capable of reproducing the rich firing patterns exhibited by biological cortical neurons from different electrophysiological classes~\cite{Nowak2003,Contreras2004}. In this study, we consider only two types of neurons, excitatory
regular spiking neurons with $a=0.02$ and $d=8$ and inhibitory fast spiking neurons with $a=0.1$ and $d=2$, and for both types, $b=0.2$
and $c=-65$~\cite{Izhikevich2003}. 

In a conductance-based synapse model~\cite{Destexhe1998,Meffin2004,Cavallari2014},
the synaptic current is related to the membrane potential: 
\begin{eqnarray} I_i(t)=  G_i^E(t)[V_E-v_i(t)] + G_i^I(t) [V_I - v_i(t)] \label{eq3}
\end{eqnarray}
where $V_E=0$ and $V_I=-80$, both in mV~{\color{revised} \cite{Cavallari2014,Tomov2014,Tomov2016,Pena2018}}, are the reversal potentials of excitatory and inhibitory synapses, and $G_i^{E}(t)$ and $G_i^{I}(t)$ are the excitatory and inhibitory conductances.
The conductance $G_i^{E (I)}$ increases by the magnitude of the synaptic weight $w_{ij}$ 
whenever a presynaptic excitatory (inhibitory) neuron $j$ fires and generates a spike at $t_{j,k}$,
otherwise it decays with a time constant $\tau_{E (I)}$~\cite{Tomov2014,Tomov2016,Pena2018}:
\begin{eqnarray}
\nonumber \frac{d G_i^{E (I)}}{dt} &=& -\frac{G_i^{E(I)}}{\tau_{E (I)}} +
 \sum_{j, {E(I)}} w_{ij} \sum_k \delta(t - t_{j,k}) \\
\Rightarrow
G_i^{E (I)}(t) &=&  \sum_{j, {E(I)} } w_{ij} \sum_{k}
e^{-(t-t_{j,k})/\tau_{E (I)}} \theta(t-t_{j,k}) \qquad  \label{eq4}
\end{eqnarray}
where $\tau_E=5$~ms and $\tau_I=6$~ms~\cite{DA2001,IE2008}. Here, $\sum_{j, E(I)}$ sums over all presynaptic excitatory (inhibitory) neurons $j$ and $\theta(t-t_0)$ is the Heaviside step function.

To facilitate theoretical analysis,
we focus on a simple directed random network (network A)  with the same excitatory and inhibitory incoming degrees ($k_{\rm in}^+=8$ and $k_{\rm in}^-=2$) for each neuron and constant excitatory and inhibitory synaptic strength, i.e.,
$w_{ij} = g_E$ or $g_I$ for neuron $j$ being excitatory or inhibitory.  Network A has $N=1000$ neurons with $N_E=800$ excitatory and $N_I=200$ inhibitory neurons, following the ratio commonly used in the study of neuronal networks~\cite{Brunel2000}. To test the robustness of our results,
we study two additional networks, {\color{new} denoted by networks B and C. N}etwork B has the same network structure as network A and the excitatory and inhibitory synaptic weights are taken from uniform distributions with mean values $g_E$ and $g_I$ with a width of 0.08, i.e. 
$[g_E-0.04,g_E+0.04]$ and $[g_I-0.04,g_I+0.04]$.

{\color{new} N}etwork C is a directed network of 4095 neurons, with $2962$ excitatory and $1133$ inhibitory neurons, reconstructed from multi-electrode array data measured from a neuronal culture~\cite{RevealPRE2022}, 
using the method of directed network reconstruction proposed by Ching and Tam~\cite{ChingTamPRE2017}, with constant excitatory and inhibitory synaptic strengths $g_E$ and $g_I$. 
{\color{revised} The excitatory and inhibitory incoming degrees of network C are non-uniform, each ranging from a few to about 100 (see Appendix).}
The degree distributions of network C are qualitatively different from those of network A. 
Network A is homogeneous in the {\color{new} excitatory and inhibitory} incoming degree by construction while network C is heterogeneous with {\color{new} bimodal  excitatory and inhibitory  incoming degree distributions. T}he outgoing degree distribution is 
long-tailed for network C and binomial for the random network A~(see Appendix).

{\color{new} We study the dynamics of these three networks as a function of $g_E$ and $g_I$ with $\alpha$ kept fixed. We have employed the first-order Euler-Maruyama and the weak second-order Runge-Kutta methods~\cite{sde} to integrate the stochastic differential equations~(\ref{SDE1}) and (\ref{SDE2}) together with equations~(\ref{eq3}) and (\ref{eq4}) with a time step $dt=0.001$~ms for networks A and B and $dt=0.005$~ms for the larger network C. The initial values of $v_i$ and $u_i$  are set to be $-70$ and $-14$, respectively. We set $\alpha=3$ for networks A and C and $\alpha=5$ for network B.} Results reported below are for network A unless otherwise stated. Similar results are found in networks B and C and will be discussed in the Appendix.

\section{Numerical Results}\label{result}

\subsection{Different dynamical states}
\label{states}

Three distinct dynamical states are found.
At a fixed value of $g_I$, including $g_I=0$ where inhibition is absent, the dynamics changes from state I of irregular and independent spiking to state II of coherent bursting and finally to state III 
of incoherent fast spiking as $g_E$ is increased. 
In state I, neurons fire irregularly, triggered essentially by $I_{\rm noise}$ and approximately independently of one another.
{\color{new} The} raster plot, which shows the times at which the spikes occur, 
thus consists of scattered random dots (see Fig.~\ref{fig1}a). 
In state II, neurons fire in bursts displaying relatively fast spiking that are separated by intervals of quiescence. There is a high coherence in the bursting dynamics of the neurons across 
the whole network but the spikes within a burst from individual neurons are not synchronized. 
The raster plot consists of stripes of densely distributed dots, corresponding to the bursts~(see Fig.~\ref{fig1}b).
In state~III, neurons exhibit fast spiking and some might still fire in bursts but the spiking or bursting is incoherent across the whole network and 
the raster plot consists of densely distributed dots that spread over time~(see Fig.~\ref{fig1}c). 

\begin{figure}[htbp]
\centering
\includegraphics[width=3.5in]{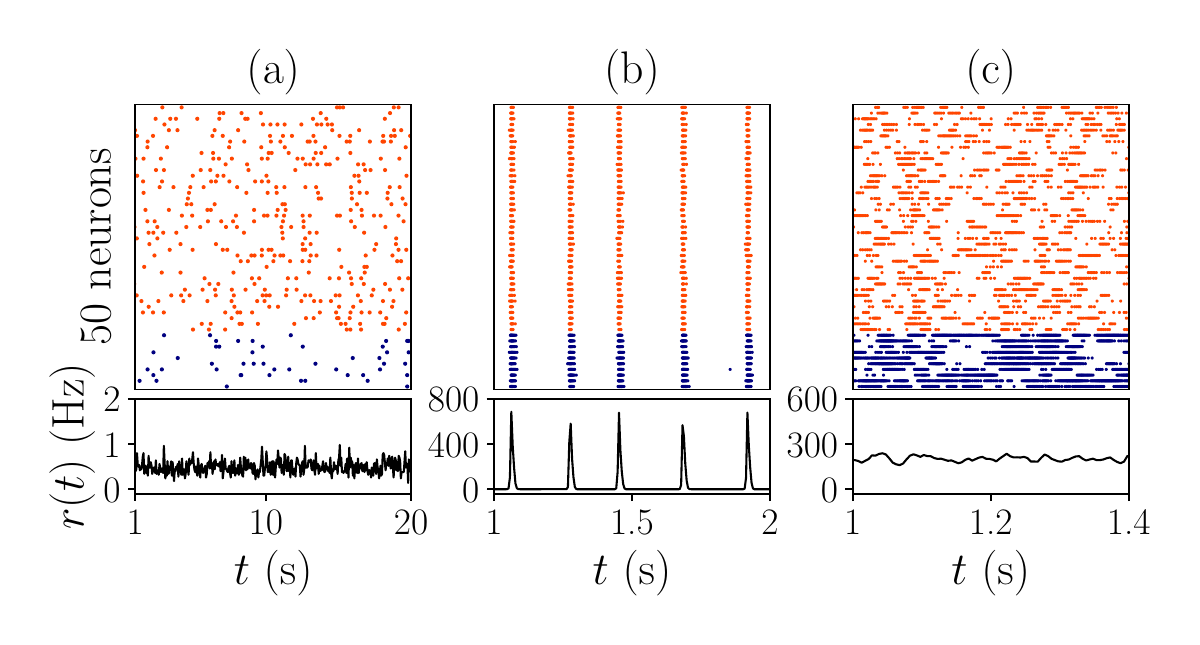}
\caption{Raster plots for 10 inhibitory (blue) and 40 excitatory (red) neurons randomly chosen and the population averaged time-binned firing rate $r(t)$ for $(g_E,g_I)$ equals to (a) $(0.04,0.2)$ in state I, (b) $(0.2,0.2)$ in state II, and (c) 
$(0.6,0.2)$ in state III. The width $w$ of the time bin used in calculating $r(t)$ is 50~ms in (a) and 5~ms in (b) and (c). }
                \label{fig1}
\end{figure}

Denote the time-binned firing rate of each neuron by $r_i(t)$, $i=1, 2, \ldots, N$. {\color{new} We calculate $r_i(t)$ by
\begin{equation}
r_i(t_n(w))=\frac{m_i(n)}{w},  \qquad n=1, 2, \ldots, N_{bin}
\label{ri}
\end{equation}
where $t_n(w)=nw-w/2$ is the the mid-point of the $n$th time bin, $w$ is the width of the time bin, $m_i(n)$ is the number of spikes of neuron $i$ occurring in the $n$th time bin, and  $N_{bin} w$ is the total observation time.} The population averaged time-binned firing rate is
\begin{equation} 
r(t) = \frac{1}{N} \sum_{i=1}^N r_i(t)
\label{r}
\end{equation}
 {\color{revised} 
{\color{new} As shown in Fig.~\ref{fig1}, $r(t)$ is approximately time-independent in states I and III whereas in} state II,  the coherent bursting
gives rise to an oscillatory $r(t)$.
The extent of temporal fluctuation of the spiking activity of each neuron is characterized by the variance of $r_i(t)$:
\begin{equation}
{\color{new} \sigma^2_i(w) = \langle [r_i(t_n(w))]^2 \rangle_t - \langle r_i(t_n(w)) \rangle_t^2 }
\end{equation}
where $\langle \cdots \rangle_t = \sum_{n=1}^{N_{bin}} ( \cdots )/N_{bin}$.
Similarly, the extent of temporal fluctuation of the population averaged spiking activity is characterized by the variance of $r(t)$:
\begin{equation} 
{\color{new} \sigma^2_{whole}(w) =\langle [r(t_n(w))]^2 \rangle_t - \langle r(t_n(w)) \rangle_t^2 }
\end{equation}

To measure the degree of coherence or synchrony of the dynamics of neurons across the whole network, we define a coherence parameter $C$
\begin{equation}
C \equiv  \left\{ \frac{\sigma^2_{whole}(w)}{ \sum_{i=1}^{N} \sigma^2_i(w)/N} \right\} \bigg |_{w=w_0}
\end{equation}
evaluated at $w_0=32$~ms, which is chosen to be about the average duration of the bursts (average width of the stripes in the raster plot) in state II. The value of $C$ ranges from 0 to 1, and for the case of total coherence with $r(t) = r_i(t)$, $C=1$. 
Our definition of the coherence parameter modifies an earlier definition~\cite{VH2001} with the replacement of the membrane potentials of neurons by their time-binned firing rates for the ease of computation. 
{\color{revised} We study the dependence of $C$ for $0 \le g_E \le 1$ at 8 different values of $g_I$ between $0$ and $8$.}
At a fixed value of $g_I$, $C$ increases from about $10^{-3}$ in state I to around 0.1-1 in state II and then decreases to about $10^{-3}$ in state III when $g_E$ is increased, as shown in Fig.~\ref{nfig2}. We use a threshold value of $C=0.03$ to mark the boundaries between the states. 
 The phase diagram obtained is shown in Fig.~\ref{nfig3}. It is clearly seen that the system can be in different states even when $g_I/g_E$ is the same. 
At a fixed value of $g_E$ in state II, $C$ decreases when $g_I$ is increased as the fraction of neurons participating in the network bursts decreases. 
The fraction of participation also varies from burst to burst leading to a variation in heights, widths and shapes of the peaks 
in the oscillatory $r(t)$~(see Fig.~\ref{nfig4}). 
{\color{new} The phase diagram obtained for network B is similar to that of network A. For network C,  states I and II can   
be clearly seen but heterogeneity of the network causes larger variations in the raster plot patterns in state III, and $r(t)$ is also weakly dependent on time in state III. Details of the results for networks B and C are presented in the Appendix.}

\begin{figure}[htbp]
\centering
\includegraphics[width=3in]{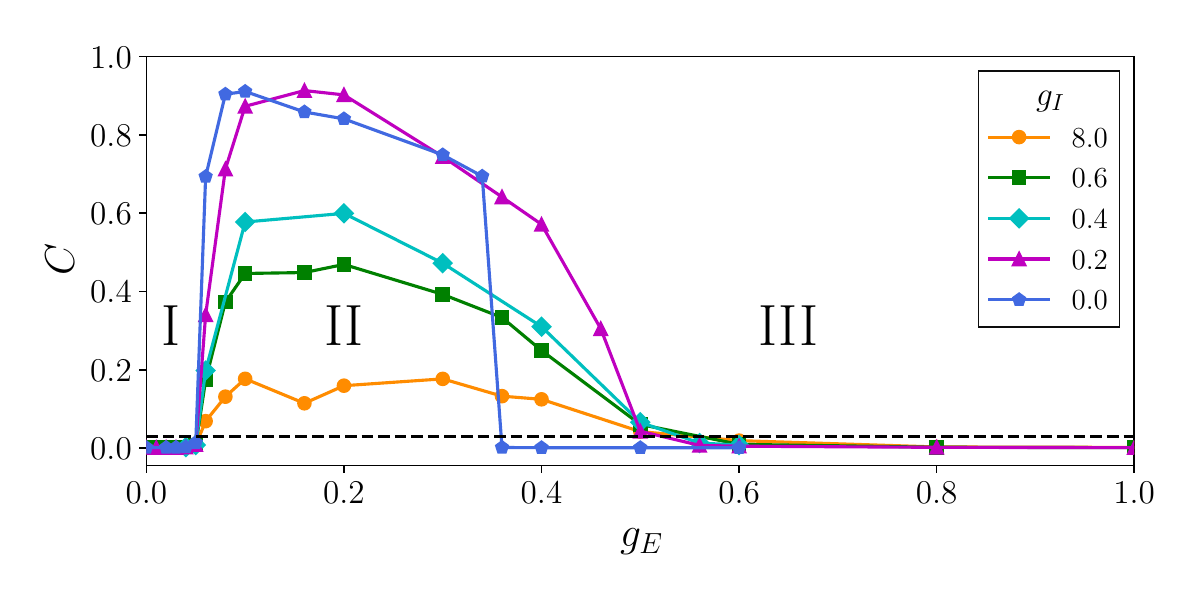}
\caption{Dependence of the coherence parameter $C$ on $g_E$ at different values of $g_I$. The dashed line corresponds to $C=0.03$, the value used to mark the boundaries between states I and II and between states II and III.}
                \label{nfig2}
\end{figure}

\begin{figure}[htbp]
\centering
\includegraphics[width=3in]{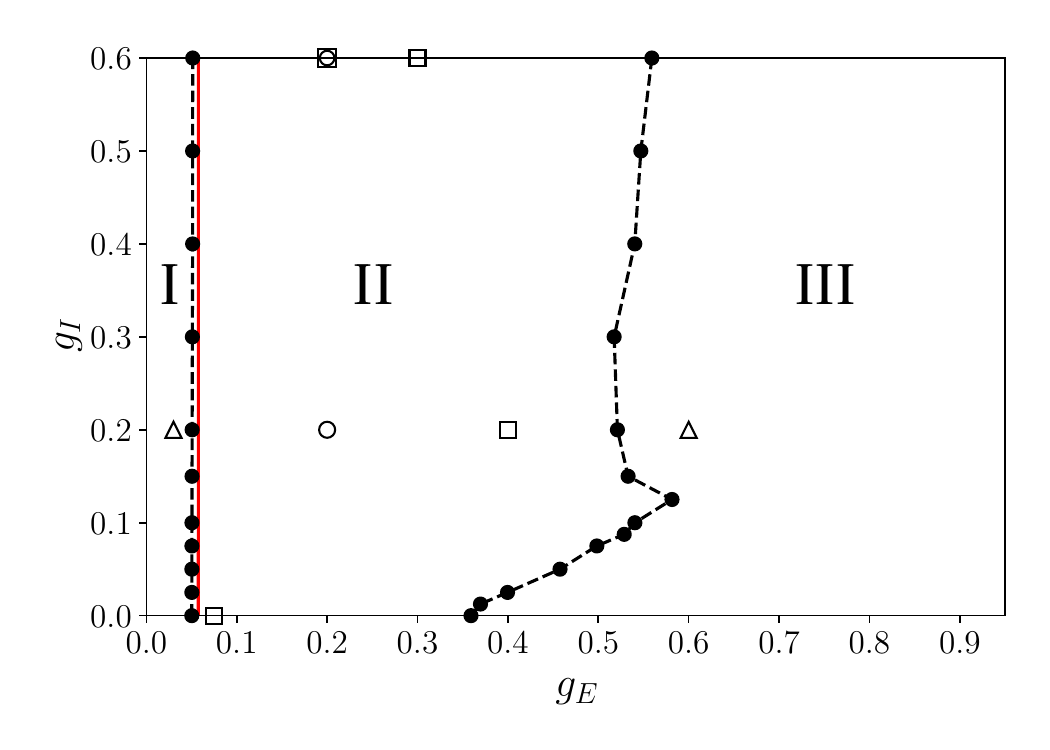}
\caption{Different dynamical states in the $g_E$-$g_I$ parameter space at $\alpha=3$. 
The dots correspond to the points of $(g_E,g_I)$ at which $C=0.03$~(see Fig.~\ref{nfig2}) and the dashed lines are linear interpolations joining the dots. The solid red line is $g_E^*(g_I)$, the theoretical upper bound of the threshold value for which state I becomes unstable~(see Sec.~\ref{cond}). {\color{new} The points $(g_E,g_I)$ in state II with $g_I \le 0.6$ used in Fig.~\ref{nfig4} and Fig.~\ref{fig6} are marked by circles and squares, respectively, and those in states I and III used in Figs.~\ref{fig4} and \ref{fig5} are marked by triangles.}}             
 \label{nfig3}
\end{figure}

\begin{figure}[htbp]
\centering
\includegraphics[width=2.7in]{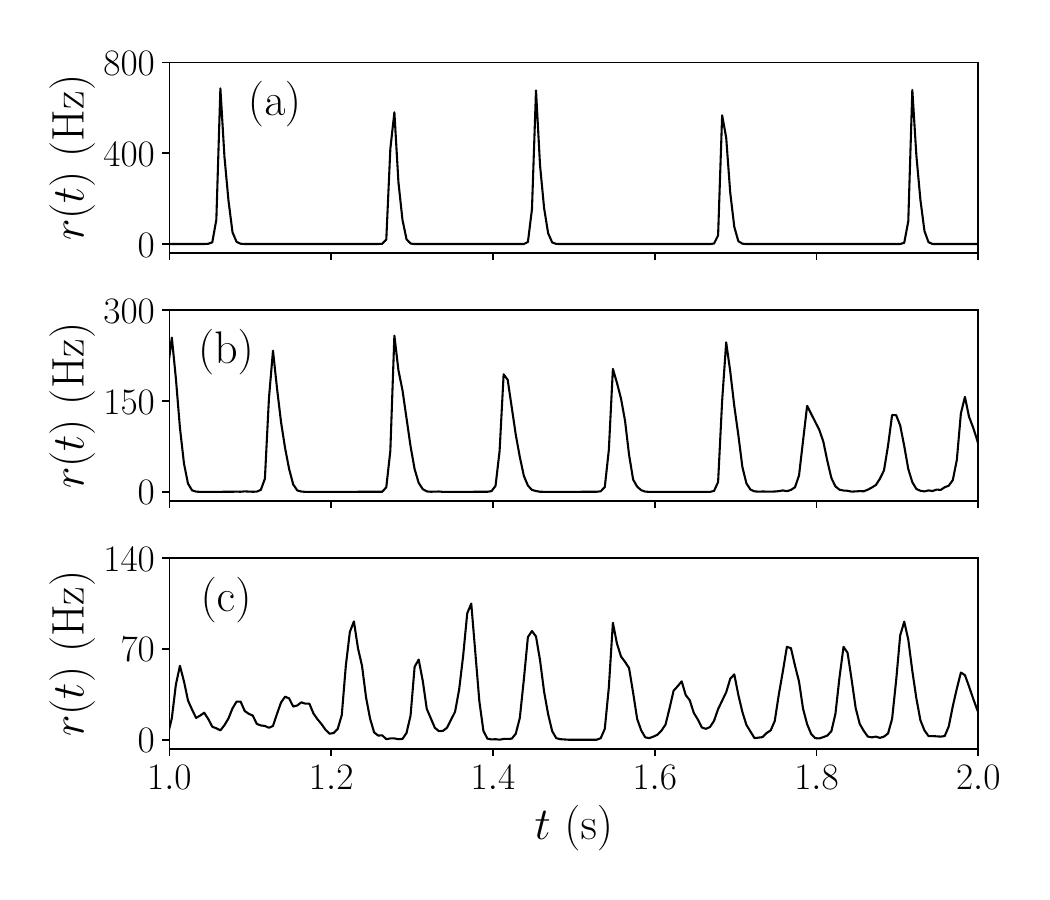}
\caption{Population averaged time-binned firing rate $r(t)$ for (a) $(g_E,g_I)=(0.2,0.2)$, (b) $(g_E,g_I)=(0.2,0.6)$, and (c) 
$(g_E,g_I)=(0.2,8)$ in state II. The width $w$ of the time bin used in calculating $r(t)$ is 5~ms.}
                \label{nfig4}
\end{figure}

{\color{revised} The seminal study of networks of excitatory and inhibitory leaky integrate-and-fire neurons with transmission delay by Brunel~\cite{Brunel2000} has revealed a rich collection of dynamical states. {\color{new} These states} are classified as asynchronous or synchronous according to whether the global activity, as measured by the population averaged time binned firing rate $r(t)$, is time-independent or oscillatory; and regular or irregular
depending on whether the individual neurons have regular or irregular firing activity. 
A later study~\cite{Ostojic2014} showed that there are two different types of asynchronous irregular states in such networks, the classical asynchronous state at 
weak excitatory coupling in which neurons fire irregularly at constant rates, which was identified by Brunel~\cite{Brunel2000}, and 
the heterogeneous asynchronous state for strong excitatory coupling in which the firing rates of individual neurons fluctuate strongly in time and across neurons. 
Both types of asynchronous irregular states occur when inhibition dominates excitation. Following these classifications, 
states I and III correspond to the classical and heterogeneous asynchronous irregular states, respectively, and state II corresponds to a synchronous irregular state. Since the Izhikevich model used in our study has spike frequency adaptation that is missing in the leaky-integrate-and-fire neurons, we shall see in Sec.~\ref{avalanches} that avalanches in state II have different features from those reported~\cite{PRE2017} in the synchronous irregular state found in Brunel's network~\cite{Brunel2000}.}

\subsection{Avalanches}
\label{avalanches}

In practice, avalanches are extracted from measured signals using an operational definition put forward by Beggs and Plenz~\cite{BeggsPlenz2003}. 
In a similar manner, we use the spiking activities of the neurons to define avalanches. The time series of the spiking activities of the whole network are partitioned into time bins with a width of $\Delta t$. 
The time-binned spiking activity of the whole network can be treated as the time-binned activity  of an effective neuron, which spikes in a certain time bin whenever any individual neuron spikes in that time bin.
A null event of the effective neuron is identified in a certain time bin if there are no spiking activities from all neurons in that time bin. An avalanche is defined as a sequence of spiking activities in consecutive time bins between 
two null events of the effective neuron. The duration $T$ of an avalanche is 
\begin{equation}
T = n \Delta t, 
\label{T}
\end{equation}
where $n$ is the number of time bins in its sequence of activity and the size $S$ of an avalanche is the total number of spikes of all neurons occurring in the duration of the avalanche.
The distributions of duration and size of the avalanches, $P(T)$ and $P(S)$, have similar forms in states I and III. As can be seen in Figs.~\ref{fig4} and \ref{fig5},  
$P(T)$ are well described by an exponential distribution
\begin{equation} 
P(T=n\Delta t)=(e^{\lambda}-1) e^{-n \lambda}
\label{PT_I_III}
\end{equation}
and the value of $\lambda$ decreases with $\Delta t$. {\color{new} The size distribution $P(S)$ is not an exponential but is close to one for large $S$.}
If $P(S)$ is an exponential function, plotting $P(S=m+1)/P(S=m)$ against $m$ should give a horizontal line. However, a deviation from a horizontal line is seen at small $S$ in such plots in the insets of Figs.~\ref{fig4}b and \ref{fig5}b.}

\begin{figure}[htbp]
\centering
\includegraphics[width=2.7in]{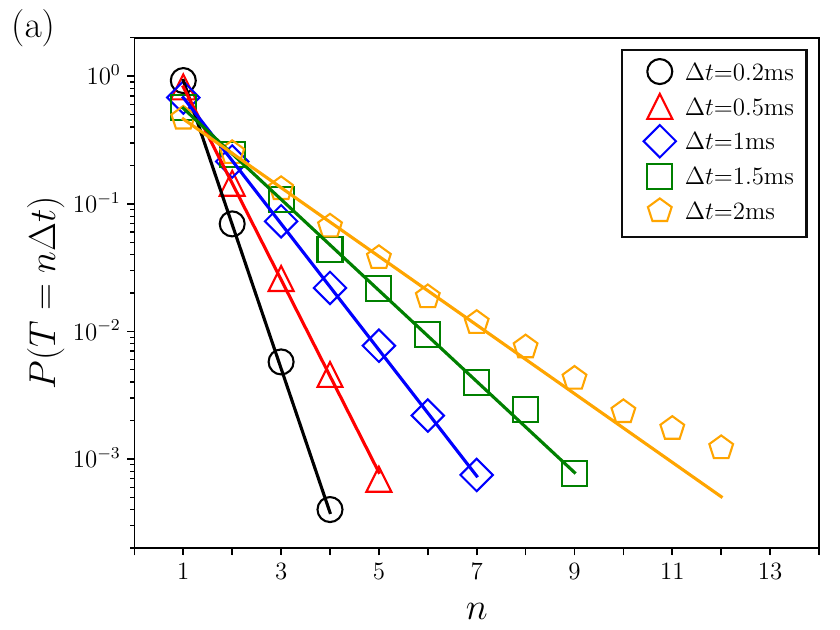}
\includegraphics[width=2.7in]{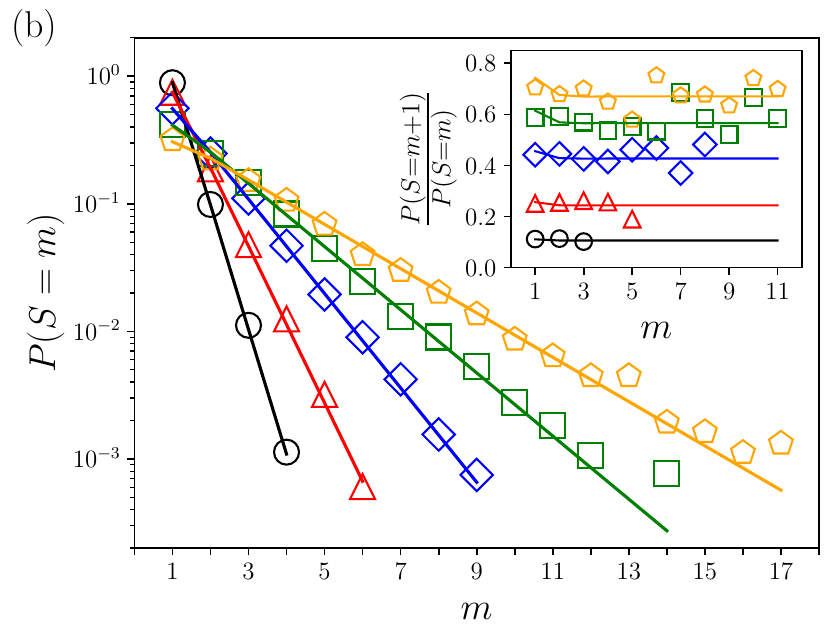}
\caption{$P(T)$ and $P(S)$ for $(g_E,g_I)=(0.03,0.2)$ in state~I for different values of $\Delta t$. The inset in (b) shows $P(S=m+1)/P(S=m)$ versus $m$. The solid lines are analytical results~Eq.~(\ref{PT_homo}) and (\ref{PS_homo}).}
                \label{fig4}
\end{figure}

\begin{figure}[htbp]
\centering
\includegraphics[width=2.7in]{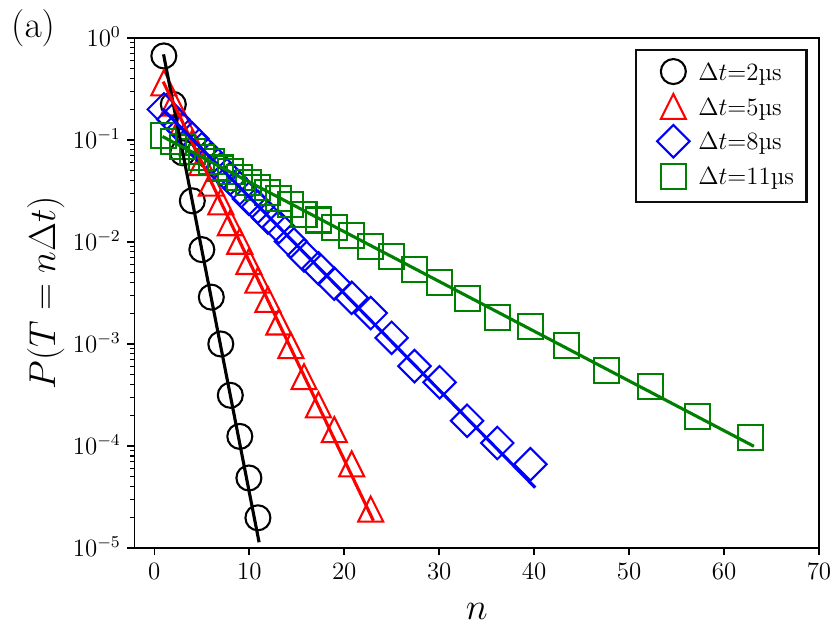}
\includegraphics[width=2.7in]{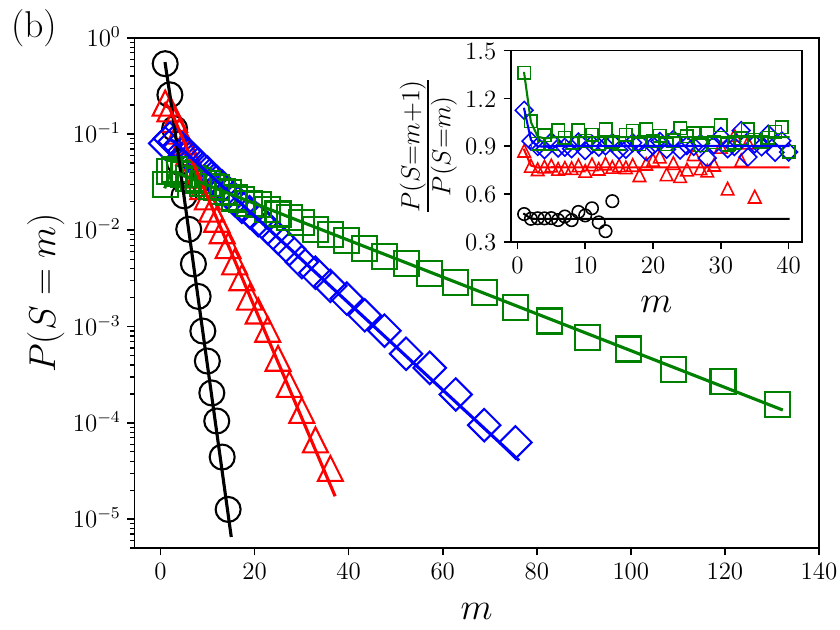}
\caption{Similar plots as in Fig.~\ref{fig4} for ${\color{new}(}g_E,g_I)=(0.6,0.2)$ in state III.}
                \label{fig5}
\end{figure}

{\color{new} 
In state I, neurons fire approximately independently of one another and the distributions of inter-spike interval of individual neurons are well described by exponential distributions as for  an isolated neuron subjected to $I_{\rm noise}$. Thus we approximate the spiking dynamics of individual neurons in this state as $N$ independent Poisson processes at constant rates given by their firing rates. The firing rate of each neuron is given by the number of spikes of that neuron divided by the total observation time. The firing rates of individual excitatory neurons are close to the average firing rate $r_E$ of all excitatory neurons. Similarly, the firing rates of individual inhibitory neurons are close to the average firing rate $r_I$ of all inhibitory neurons.} The dynamics of the whole network is  the sum of these approximate independent homogeneous Poisson processes, which is itself also approximately a homogeneous Poisson process at a constant rate $R_0=N_E r_E +N_I r_I$.  {\color{new} Using the analytical results derived for $P(T)$ and $P(S)$ in a homogeneous Poisson process~\cite{PS2018}, we obtain}
\begin{equation}
 P(T = n \Delta t) = (e^{\lambda_t} - 1) e^{-n \lambda_t} \label{PT_homo}
\end{equation}
with
\begin{equation} \lambda_t = -\ln(1 - e^{-R_0 \Delta t}) \label{lambdaTheo}
\end{equation} and
 \begin{eqnarray}
\nonumber  \hspace{-0.2cm} && P(S = m) \\ \hspace{-0.2cm} &=&
 \frac{(R_0 \Delta t)^m}{m! (e^{R_0 \Delta t} - 1)} \sum_{n = 1}^m e^{- n R_0 \Delta t} \sum_{k = 0}^n (-1)^k \binom{n}{k} (n - k)^m \label{PS_homo} \qquad
\end{eqnarray}
Equation (\ref{PT_homo}) confirms that $P(T)$ is an exponential distribution.
As discussed in~\cite{PS2018}, Eq.~(\ref{PS_homo}) shows that $P(S)$ does not equal to an exponential and does not decrease monotonically 
but can be approximated by an exponential for large $S$. 
These analytical results are in good agreement with the numerical results shown in Fig.~\ref{fig4}. 
The theoretical result, Eq.~(\ref{lambdaTheo}), for the dependence of $\lambda_t$ on $\Delta t$ is also confirmed by the numerical results~(see Fig.~\ref{fig8}).

\begin{figure}[htbp]
\centering
\includegraphics[width=2.8in]{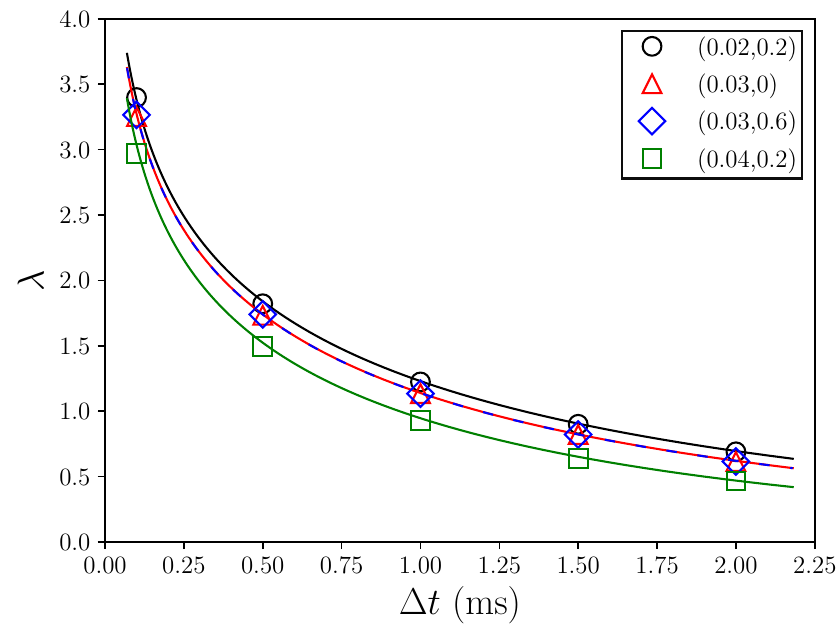}
\caption{Dependence of the measured $\lambda$ for $P(T)$ on $\Delta t$ for different values of $(g_E, g_I)$ in state I {\color{revised} [$(g_E,g_I)$=(0.02,0.2) (circles), (0.03,0) (triangles, (0.03,0.6) (diamonds) and (0.04,0.2) (squares)].
The solid lines are the theoretical result Eq.~(\ref{lambdaTheo}) calculated using $r_E$ and $r_I$ measured for different $(g_E,g_I)$ and are in good agreement with the numerical results.} }
            \label{fig8}
\end{figure}

In state III, neurons fire at heterogeneous time-dependent firing rates and {\color{new} the distribution of inter-spike intervals of individual neurons deviate from an exponential distribution, showing that} the spiking dynamics of each neuron cannot be approximated as a homogeneous Poisson process.  But the neurons fire incoherently and {\color{new} these incoherent dynamics averaged out to give an approximately time-independent $r(t)$, suggesting that the collective dynamics of state III could} be approximated by a  
homogeneous Poisson process of a constant rate $R_0$, given by the total number of spikes of all the neurons divided by the total observation time.}
Indeed, the analytical results, Eqs.~(\ref{PT_homo}) and (\ref{PS_homo}), for $P(T)$ and $P(S)$ of a homogeneous Poisson process are in good agreement with the numerical results shown in Fig.~\ref{fig5}.

\begin{figure*}[htbp]
\centering
\includegraphics[width=5.5in]{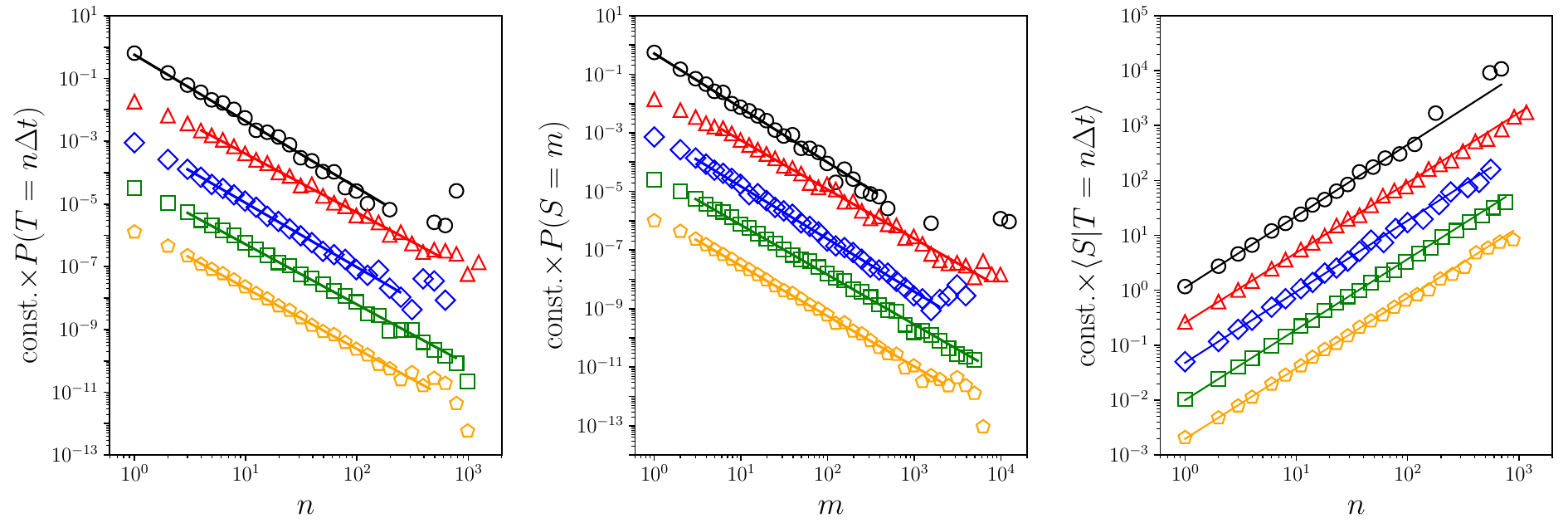}
\caption{$P(T)$, $P(S)$ and $\langle S|T\rangle$ for $(g_E, g_I) = (0.075,0), (0.4,0.2), (0.2,0.6), (0.3,0.6)$ and $(0.3,8)$, from top to bottom, in state~II for $\Delta t={\rm IEI}_{\rm ave}$ except for $(0.3,8)$ where $\Delta t=2 {\rm IEI}_{\rm ave}$. {\color{new}The symbols are numerical results while the solid lines are power-law fits of $P(T)$ and $P(S)$ using maximum likelihood estimation and least-squared fits of $\langle S|T \rangle$ in log-log scale.}
For clarity, all the symbols and solid lines, except for the topmost ones, are shifted downwards by multiplying arbitrary constants.}
            \label{fig6}
\end{figure*}

\begin{figure}[htbp]
\centering
\includegraphics[width=2.8in]{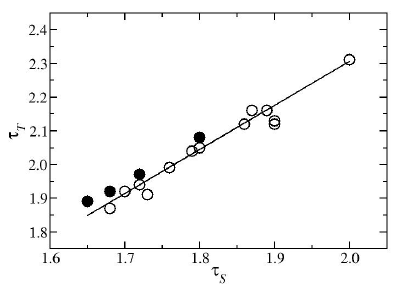}
\caption{Relation between $\tau_T$ and $\tau_S$ for all the cases in state II studied as shown in Table~\ref{table1}. The filled symbols correspond to $(g_E,g_I)=(0.3, 0.2)$ with different values of $\Delta t$. The solid line is the least-squares fit $(\tau_T-1)/(\tau_S-1)=K$ {\color{new} and the least-squares estimate of the constant $K$ is $K=1.3$.}}
            \label{fig7}
\end{figure}

At each pair of fixed values of $(g_E,g_I)$ in state II,  $P(T)$ and $P(S)$ are found to be well described by power laws, and $\langle S|T \rangle$
has a power-law dependence on $T$ for a certain range of $\Delta t$. Within this range,
the exponents $\tau_T$ and $\tau_S$ vary with $\Delta t$ as found in experiments~\cite{BeggsPlenz2003,Petermann2009} but $\gamma$ remains approximately the same.
The power-law exponents for $\tau_T$ and $\tau_S$ are estimated by the maximum likelihood estimator of a discrete power law distribution and the hypothesis that the data are described by a power law is validated by the 
goodness-of-fit test using the Kolmogorov-Smirnov statistic with a  requirement of $p$-value greater than $0.1$~\cite{Siam2009}. A common choice of  $\Delta t$ is the average inter-event interval, IEI$_{\rm ave}$, 
of the effective neuron~\cite{BeggsPlenz2003}. In Fig.~\ref{fig6}, we show $P(T$), $P(S)$ and $\langle S | T \rangle$ for {\color{new} $\Delta t={\rm IEI}_{\rm ave}$, and when IEI$_{\rm ave}$ does not fall within the range of $\Delta t$ for which power-law distributions are observed, we choose instead a $\Delta t$ that is around the middle of the range.} 
Our results for $\tau_T$, $\tau_S$ and $\gamma$ for all the cases studied in network A are presented in Table~\ref{table1} {\color{new} and they satisfy approximately the scaling relation Eq.~(\ref{relation}). Hence, our results differ from the values $\tau_T=2.11$, $\tau_S=1.42$ and $\gamma=1.5$ found
in the synchronous irregular state in a network of leaky-integrate-and-fire neurons~\cite{PRE2017}, which do not satisfy Eq.~(\ref{relation}).}
Furthermore, as shown in Fig.~\ref{fig7}, our results give $(\tau_T-1)/(\tau_S-1) \approx 1.3$, which is very close to the value $1.28$ observed in experiments~\cite{Fontenele2019}.
All these results are found throughout the entire state II, and 
also in state II of networks B and C (see Appendix). 

\begin{table}[htbp]
    \centering
    \begin{tabular}{|c|c|c|c|c|c|c|} \hline 
    $g_E$&  $g_I$&  $\tau_T$&  $\tau_S$ & $\displaystyle \frac{(\tau_T-1)}{(\tau_S-1)}$ &  $\gamma$& $\Delta t \text{ (ms)}$\\ \hline 
         0.075&  0&  2.12&  1.86& 1.30 & 1.30& 0.03\\ \hline 
         0.1&  0&  1.94&  1.72&  1.31& 1.28& 0.005 (0.021)\\ \hline 
         0.2&  0&  2.16&  1.87& 1.33&  1.23& 0.014\\ \hline 
         0.075&  0.2&  2.04&  1.79& 1.32&  1.29& 0.04 (0.147)\\ \hline 
         0.1&  0.2&  1.91&  1.73&  1.25& 1.27& 0.02 (0.066)\\ \hline 
         0.2&  0.2&  1.99&  1.76&  1.30 & 1.29& 0.015 (0.037)\\ \hline 
         0.3&  0.2&  2.08&  1.80& 1.35&  1.30& 0.027\\ 
          0.3 &  0.2 &  1.89 & 1.65 & 1.37 & 1.27 & 0.04 (0.027) \\
          0.3 & 0.2 & 1.92 & 1.68 & 1.35 & 1.27 & 0.05 (0.027) \\
          0.3  & 0.2 &  1.97 & 1.72 & 1.35&  1.27  & 0.06 (0.027) \\ \hline
         0.4&  0.2&  1.87&  1.68&  1.28 & 1.27& 0.019\\ \hline
         0.075& 0.6& 2.12& 1.90& 1.24& 1.29&0.1 (0.195)\\ \hline
         0.1& 0.6& 2.31& 2.00& 1.31 & 1.28&0.115\\ \hline
         0.2& 0.6& 2.05& 1.80& 1.31& 1.29&0.036\\ \hline
         0.3& 0.6& 1.92& 1.70& 1.31& 1.28&0.02\\ \hline
         0.1& 8& 2.13& 1.90& 1.26 & 1.31&0.16 (0.139)\\ \hline
         0.2& 8& 2.16& 1.89&1.30&  1.25&0.06 (0.043)\\ \hline
         0.3& 8& 1.94& 1.72& 1.31 & 1.29&0.04 (0.02)\\\hline
   \end{tabular}
    \caption{The exponents $\tau_T$, $\tau_S$ and $\gamma$ for the different cases studied. 
    When $\Delta t \ne \text{IEI}_\text{ave}$, the values of $\text{IEI}_\text{ave}$ are indicated in parentheses. The dependence of the exponents on $\Delta t$ is illustrated for 
    $(g_E,g_I)=(0.3,0.2)$. }
    \label{table1}
\end{table}

{\color{new}
\section{Understanding Avalanches in State II} \label{Theory}

 \subsection{Origin of Power-law Distributed Avalanches}}

\begin{figure*}[htbp]
\centering
\includegraphics[width=6.2in]{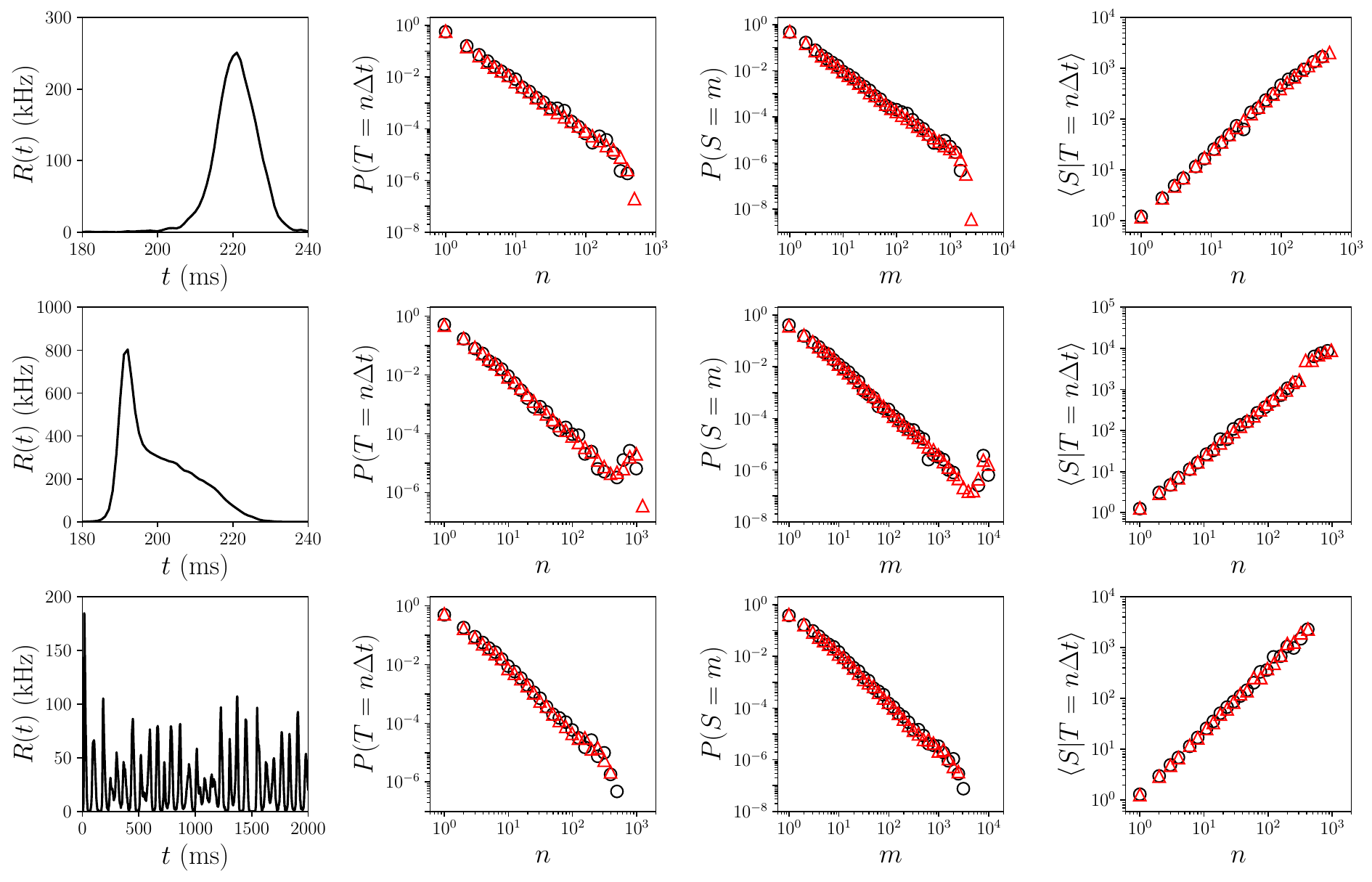}
\caption{Comparison of $P(T)$, $P(S)$ and $\langle S|T \rangle$ obtained from an inhomogeneous Poisson process with a time-varying rate $R(t)$ and 
from the neuronal network in state II. {\color{new} For $(g_E,g_I) = (0.1,0.2)$ (top) and $(g_E,g_I)=(0. 3,0.2)$ (middle), $R(t)$ for one typical peak of the oscillatory population time-binned firing rate is shown. For $(g_E,g_I)=(0.2,8)$ (bottom), $R(t)$ for the whole simulation duration of 10~s is used and 2s of which is shown.} Results from simulations of the inhomogeneous Poisson process and the neuronal network are denoted by triangles and circles, respectively. }
            \label{fig10}
\end{figure*}

Power-law distributed neuronal avalanches are found in the entire state II while exponential duration distribution and approximately exponential 
size distributions are found in states I and III. 
As the power-law distributions do not arise when the system is near a transition point, it is not likely that they are due to criticality of the system.
 {\color{new} The defining} feature of state II is coherent network bursting and it gives rise to an oscillatory population averaged time-binned firing rate $r(t)$~(see
Fig.~\ref{nfig4}). This leads us to the idea that power-law distributed neuronal avalanches are generated by the stochastic collective dynamics of an effective neuron with a firing rate given by {\color{new} the population time-binned firing rate} $R(t)=Nr(t)$.  {\color{new} We further approximate this stochastic collective dynamics of an effective neuron} by an inhomogeneous Poisson process with the time-varying rate $R(t)$. 
{\color{new} To test our idea, we  simulate an inhomogeneous Poisson process of rate $R(t)$ using $\delta t=0.001$~ms and} 
generate the number of spikes $k$ in each time interval $[t, t+\delta t]$ 
according to the Poisson distribution:
\begin{equation}
P(k)  =  \frac{[R(t) \delta t]^k }{k!} e^{-R(t) \delta t} 
\label{inhomoPoisson}
\end{equation}
To obtain a {\color{new} smooth} $R(t)$ from the numerical data of the network simulations, we calculate the time-binned firing rate $r(t)$ using a bin width of 1~ms and smooth the result using a Gaussian filter with a standard deviation of 1~ms {\color{new} and cubic spline interpolation.}
For small $g_I$, the peaks in the oscillatory $r(t)$ are similar~(see Fig.~\ref{nfig4}a) and we further approximate $R(t)$ as a repetition of one typical peak.  
To extract $R(t)$ for one typical peak, we truncate the data on both sides of the peak
when there is no spiking activity in at least 5 consecutive time bins.
For larger $g_I$, there are large variations in the heights, widths and shapes of the peaks~(see Fig.~\ref{nfig4}c) and {\color{new} we} use $R(t)$ obtained from the whole duration of the network simulation. {\color{new} For $R(t)$ approximated by a repetition of one typical peak, we repeat the simulation for the extracted peak for 4000 times. This large number of repetitions, which far exceeds the number of peaks (about 25) observed in the whole duration of the network simulation, allows us to obtain good statistics of avalanches of large size and long duration and reveal a scaling form of the conditional size distribution for a given duration~(see Sec.~\ref{scalingform}).} Using the time series of the number of spikes {\color{new} generated in the inhomogeneous Poisson process simulation,} we carry out avalanche analysis as in the network simulation.
As shown in Fig.~\ref{fig10}, $P(S)$ and $P(T)$ and $\langle S|T\rangle$ so obtained match perfectly the numerical results found in the network simulations, confirming our idea. {\color{new} This finding is important as it shows directly that} the origins of the power-law distributed avalanches, with properties described by Eqs.~(\ref{duration})-(\ref{relation}), in state II are stochasticity and coherent bursting from which the oscillatory $R(t)$ arises. There remains the natural question of what conditions are required for coherent bursting to occur and we shall address this question in Sec.~\ref{cond}. {\color{new} Our results also show that inhomogeneous Poisson processes with suitable time-varying rates, such as the oscillatory $R(t)$ found in state II, can generate power-law distributed avalanches that satisfy Eq.~(\ref{relation}) in contrast to the analytical results derived for inhomogeneous Poisson processes in the slow rate regime in~\cite{PRE2017}.}

\vspace{0.5cm}

\subsection{Scaling Form of the Conditional Size Distribution}
\label{scalingform}

\begin{figure}[htbp]
\centering
\includegraphics[width=3.5in]{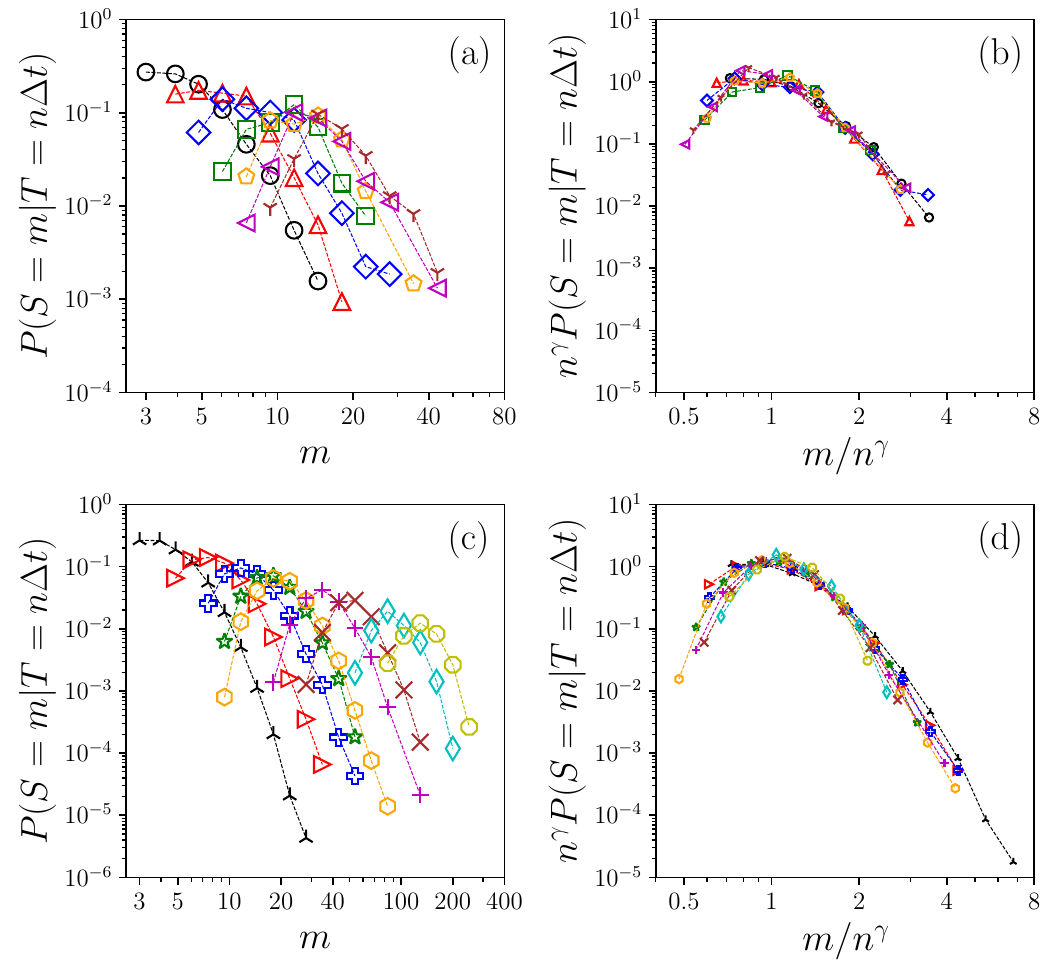}
\caption{Scaling of $P(S=m|T=n\Delta t)$ for (a)-(b) neuronal network with $n=$ 3-9 and (c)-(d) inhomogeneous Poisson process with $n$ ranging from 3 to 40 for $(g_E,g_I)=(0.3,0.2)$ in state II. $\gamma=1.3$ for (b) and 1.29 for (d).}
            \label{fig12}
\end{figure}

Further analysis of the neuronal avalanches in state II {\color{new} shows that all the data of the conditional size distribution for a given duration, denoted by $P(S|T)$, approximately collapse into one single curve when suitably rescaled. This is shown in Fig.~\ref{fig12} where $P(S|T) T^\gamma$ is plotted against $ST^{\gamma}$ for  data from both the neuronal network simulation and the inhomogeneous Poisson process simulation. The collapsed curve is denoted by $F(x)$ with $x=ST^{-\gamma}$. Such an approximate collapse of data reveals an important approximate scaling form of $P(S|T)$:}
\begin{equation}
P(S|T) {\ \color{new} \approx \ } T^{-\gamma} F(ST^{-\gamma}), \qquad \gamma > 0 \ ,
\label{scaling}
\end{equation}
{\color{new} where $F(x)$ is a continuous function that decays sufficiently fast such that $\int_0^\infty x^\beta F(x) dx$ is finite for $\beta >0$.}
The good statistics of avalanches of large sizes and long durations from the inhomogeneous Poisson process simulation enable us to visualize a broad range of values of $F(x)$ over more than 4 decades. This scaling form is clearly different from the delta function, {\color{new} $P(S|T)~\sim~\delta(S-T^\gamma)$,} assumed in~\cite{Fosque2021} and, to the best of our knowledge, has not been reported in the literature.
 
The importance of Eq.~(\ref{scaling}) is that it relates the four observed properties, Eqs.~(\ref{duration})-(\ref{relation}), of neuronal avalanches.
Specifically,  we show below that Eq.~(\ref{gamma}) follows from Eq.~(\ref{scaling}) and Eq.~(\ref{size}) and Eq.~(\ref{relation}) follow from Eq.~(\ref{duration}) and Eq.~(\ref{scaling}) when the discrete distributions are approximated by continuous ones.
First, \begin{equation} \langle S | T \rangle \approx  \int_0^\infty S P(S|T) dS {\ \color{new} \approx \ } T^\gamma \int_0^\infty x F(x) dx  
\end{equation}
and thus $\langle S| T \rangle \sim T^\gamma$, which is Eq.~(\ref{gamma}).  Second,
\begin{eqnarray} 
\nonumber P(S) &\approx& \int_0^\infty P(S | T) P(T) dT \\
& \sim&  S^{-[1+(\tau_T-1)/\gamma]} \int_0^\infty x^{(\tau_T-1)/\gamma} F(x) dx
\end{eqnarray}
thus $P(S) \sim S^{-[1+(\tau_T-1)/\gamma]}$ is power-law distributed, as described by Eq.~(\ref{size}), and $\tau_S = 1+ (\tau_T-1)/\gamma$, which is just Eq.~(\ref{relation}). 
{\color{new} Hence, the problem of explaining the four observed properties of neuronal avalanches, namely power-law duration and size distributions, power-law dependence on the duration of the conditional average avalanche size $\langle S| T \rangle$, and the scaling relation between the power-law exponents is reduced to that of explaining why neuronal avalanches satisfy Eq.~(\ref{scaling}) and have power-law duration distribution.}

{\color{revised} \section{Conditions for the Existence of Coherent Bursting}
\label{cond}

{\color{new} State I of irregular and independent spiking occurs at small values of $g_E$. A necessary condition for a new dynamical state, state II of coherent bursting, to occur is that state I becomes unstable when $g_E$ is large. We are going to show this instability of state I.}
Let the bare firing rate of an isolated excitatory or inhibitory neuron induced by the stochastic current $I_{\rm noise}$ be $r_E^{(0)}$ or $r_I^{(0)}$.
In the presence of coupling with the other neurons, a single spike from a presynaptic excitatory (inhibitory) neuron of a certain neuron, would 
generate a presynaptic current. In state I, because of the small $g_E$, this presynaptic current alone is too small to induce a spike but it changes the probability of spike initiation by the stochastic current in that neuron and thus changes its firing rate.
Denote the changes in the firing rates due to the bare firing rates as $r_E^{(1)}$ and $r_I^{(1)}$, then we have
 \begin{eqnarray}
r_E^{(1)} &=& \rho_{EE} k_{in}^+ r_E^{(0)} + \rho_{EI} k_{in}^-  r_I^{(0)} \label{rE0}\\ 
 r_I^{(1)} &=&\rho_{IE} k_{in}^+  {\color{new} r_E^{(0)}} + \rho_{II} k_{in}^- r_I^{(0)} \label{rI0}
\end{eqnarray}
where $\rho_{XY}$ is the change in firing rate of a neuron of type $X$ ($E$ or $I$) due to one single spike of a presynaptic neuron of type $Y$ ($E$ or $I$) and 
$k_{in}^{+(-)}$ is the incoming excitatory (inhibitory) degree of a neuron. 
Similarly, $r_E^{(1)}$ and $r_I^{(1)}$ induce further changes in firing rates, denoted by $ r_E^{(2)}$ and $ r_I^{(2)}$, and the process continues.
The changes in firing rates, $r_E^{(n+1)}$ and $r_I^{(n+1)}$, and the changes in firing rates $r_E^{(n)}$ and $r_I^{(n)}$ are similarly related as in Eqs.~(\ref{rE0}) and (\ref{rI0}), which can be written as,
\begin{align}
\begin{bmatrix} 
    r_E^{(n+1)} \\[1mm] 
    r_I^{(n+1)}  \\
\end{bmatrix} =   {\bf P K} \begin{bmatrix} 
    r_E^{(n)} \\[1mm]
    r_I^{(n)}  \\
\end{bmatrix} 
\end{align}
for $n=1, 2, \ldots$, where
${\bf P}$ and ${\bf K}$ are defined by
\begin{align}
{\bf P}=\begin{bmatrix} 
    \rho_{EE} & \rho_{EI} \\[1mm]
    \rho_{IE} & \rho_{II} \\
\end{bmatrix},\qquad
{\bf K}=\begin{bmatrix} 
    k_{in}^+ & 0 \\[1mm]
    0 & k_{in}^- \\
\end{bmatrix}
\end{align}
The actual firing rate of {\color{new} an excitatory (inhibitory)} neuron in the network, {\color{new} which is given by $r_E$ ($r_I$),} is equal to the sum of the bare firing rate and all these successive changes in the firing rate.  Thus $r_{E(I)}=\sum_{n=0}^{\infty} r_{E (I)}^{(n)}$ and hence
\begin{align}
\begin{bmatrix} 
    r_E \\[1mm] 
    r_I  \\
\end{bmatrix} =  \sum_{n=0}^{\infty} ({\bf P K})^n \begin{bmatrix} 
    r_E^{(0)} \\[1mm]
    r_I^{(0)}  \\
\end{bmatrix} = ({\bf I} - {\bf PK})^{-1} \begin{bmatrix} 
    r_E^{(0)} \\[1mm]
    r_I^{(0)}  
    \label{rateDep}
\end{bmatrix} \end{align} 
We measure $\rho_{EE}$, $\rho_{IE}$, $\rho_{EI}$ and $\rho_{II}$ by numerical simulations of isolated neurons subjected to the same stochastic input with one additional spike with $g_E$ or $g_I$. 
By plotting the determinant of $({\bf I} - {\bf PK})$ as a function of $g_E$ for different fixed values of $g_I$, we find that
it becomes zero at $g_E^*(g_I)$. 
 This result indicates that if the spiking dynamics remain independent and irregular, the firing rates would diverge at $g_E^*(g_I)$. The firing rates {\color{new} could} not diverge in reality; instead
the network will undergo a transition from state I to a different dynamical state in which spiking dynamics of the neurons are coupled and correlated 
when $g_E$ is raised above a certain threshold, and $g_E^*(g_I)$ is a theoretical upper bound of this threshold value.  Indeed, 
{\color{new} $g_E^*(g_I) \approx 0.057$} is slightly above the boundary between states I and II, {\color{new} which occurs at $g_E \approx 0.05$~(see Fig.~\ref{nfig3}).} In Fig.~\ref{fig9}, we verify that Eq.~(\ref{rateDep}) can describe the dependence of the firing rates well for $g_E < 0.05$, close to and below $g_E^*(g_I)$. {\color{new} At $g_E=0.05$ just before the transition from state I to state II, spiking of neurons is random and incoherent but $r(t)$ exhibits large fluctuations. Then at $g_E=0.06$ immediately after the transition, spiking dynamics of neurons becomes coherent giving rise to an oscillatory $r(t)$, similar to that shown in Fig.~\ref{fig1}b, and a significant increase in the coherence parameter $C$~(see Fig.~\ref{nfig2}).}
 
\begin{figure}[htbp]
\centering
\includegraphics[width=3in]{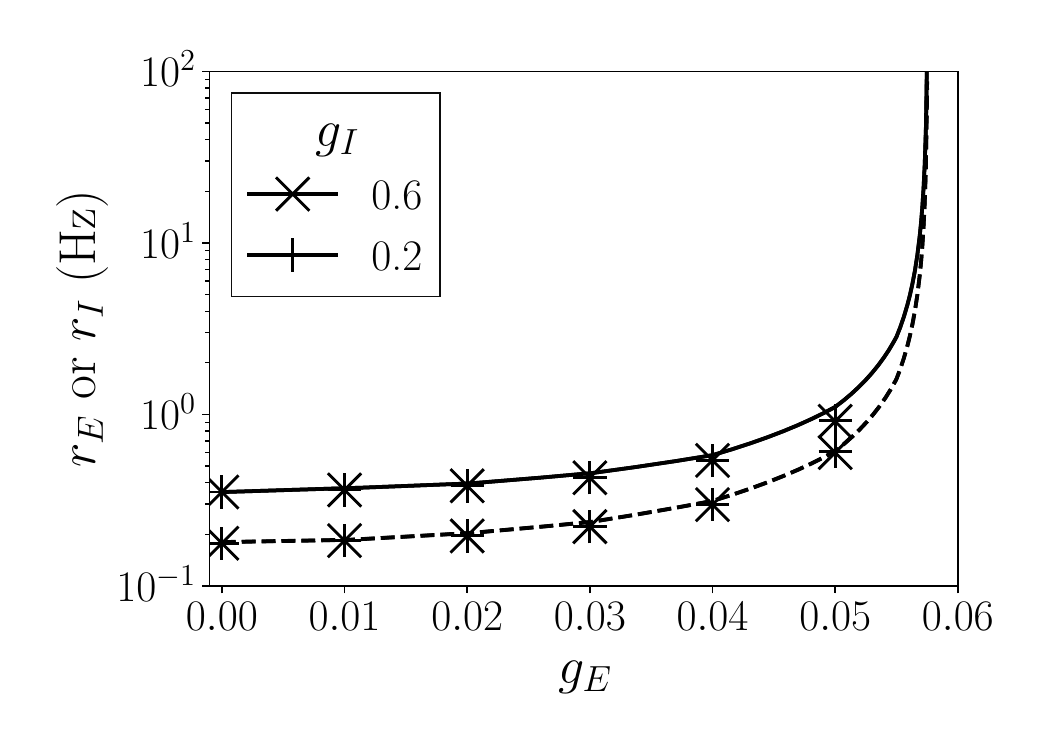}
\caption{Dependence of {\color{new} the average firing rates $r_E$ and $r_I$ of excitatory and inhibitory neurons} on $g_E$ for two values of $g_I$. The numerical values (symbols) are in good agreement with the theoretical results (solid lines) given by Eq.~(\ref{rateDep}). }
            \label{fig9}
\end{figure}

{\color{new} We shall see that strong excitation alone cannot lead to coherent bursting.}
Earlier studies suggest that coherent bursting could result from the cooperation between strong excitation and sufficiently strong adaptation in networks of adaptive neurons~\cite{VH2001,Ferguson2015,Fardet2018}.
We investigate the role of adaptation explicitly using numerical simulations with varying adaption strength by replacing $d$ in 
Eq.~(\ref{adapt}) with $\kappa d$ and varying {\color{new} the parameter $\kappa$, which measures the strength of adaptation:}
\begin{eqnarray}
\begin{cases}v_i \rightarrow c\\
u_i \rightarrow u_i + \kappa d \end{cases}
\label{adaptnew}
\end{eqnarray}
In Fig.~{\color{new} \ref{fig11}}, we show the phase diagram with the boundaries between the states in the $g_E$-$\kappa$ parameter space for a few values of $g_I$ that summarizes the effect of adaptation.
At a fixed value of $g_I$, the boundary between states II and III is given by a curve $\kappa^*(g_E,g_I)>0$ in the $\kappa$-$g_E$ plane, above which state II exists and below which state III exists. 
Specifically, if a network is in state II at $\kappa =1$, it would undergo transition to state III as $\kappa$ is decreased below $\kappa^*<1$ with $g_E$ and $g_I$ held fixed. 
Similarly, a network in state III at $\kappa=1$ would undergo transition to state II as $\kappa$ is increased above $\kappa^*>1$ with $g_E$ and $g_I$ held fixed. 
Thus, at a fixed adaptation strength with a fixed value of $\kappa$ greater than $ \kappa^*$, state II exists in a range of $g_E$  given approximately by $g_E^*(g_I) \le g_E \le \hat{g}_E(g_I,\kappa)$, where $\hat{g}_E(g_I,\kappa)$ is obtained from $\kappa^*(g_E,g_I)=\kappa$. 
{\color{revised} We note the important result of $\hat{g}_E(g_I,\kappa) \to g_E^*(g_I)$ as $\kappa \to 0$, which implies that in the absence of adaptation, state II would not exist and the network would undergo transition from state I directly to state III when $g_E$ is increased, for all values of $g_I$. 
Hence, state II of coherent bursting exists when excitation is sufficiently strong and balanced by adaptation.}

\begin{figure}[htbp]
\centering
\includegraphics[width=2.8in]{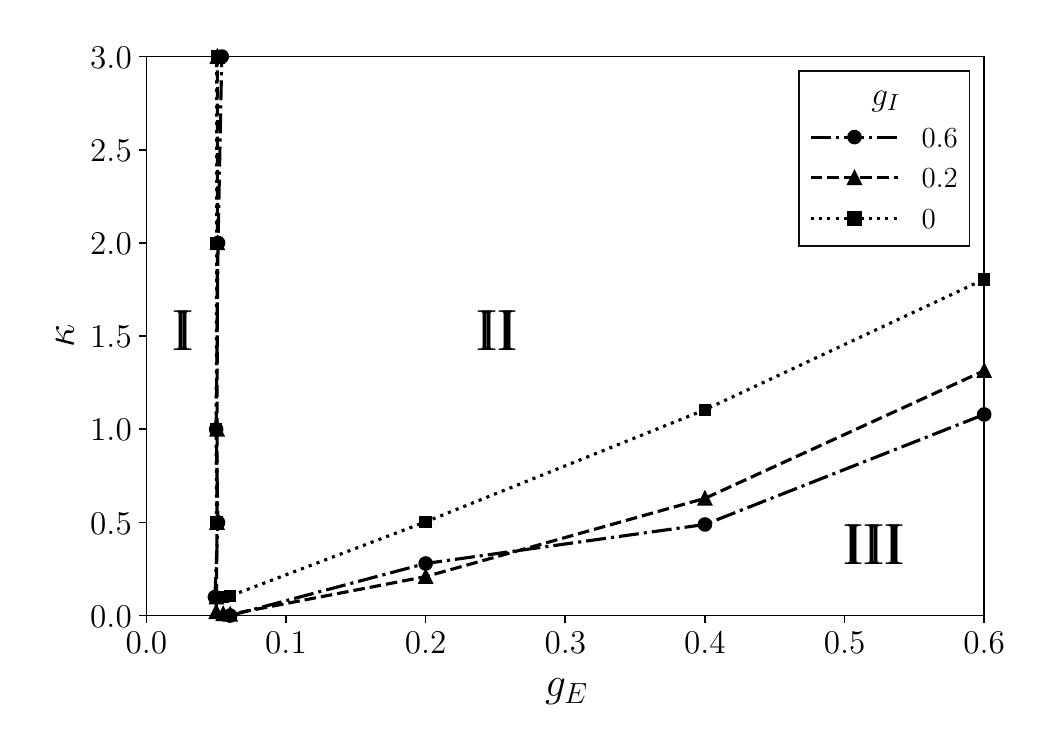}
\caption{Boundaries between states I, II, and III in the $\kappa$-$g_E$ parameter space for $g_I=0$ (dotted line), $g_I=0.2$~(dashed line) and $g_I=0.6$~(dot-dashed line).
The boundary between state II and state III defines the curve $\kappa^*(g_E,g_I)$.}
            \label{fig11}
\end{figure}

\section{Discussion and Conclusion}

Spontaneous brain activity displays rich spatial and temporal dynamical structures such as oscillations and power-law distributed neuronal avalanches.
Understanding its underlying mechanism and functional significance is a fundamental challenge in neuroscience.
The observed power-law distributed neuronal avalanches have been interpreted as evidence supporting the hypothesis that the brain is operating at the edge of a continuous phase transition, and have thus attracted much attention.
However, this critical brain hypothesis {\color{revised} and, in particular, whether the observed dynamical features are signatures of criticality in brain dynamics} remain controversial. 
Moreover, there is not yet a study showing {\color{new} that both} oscillations and power-law distributed neuronal avalanches with the observed features,~Eqs.~(\ref{duration})-(\ref{relation}) and $(\tau_T-1)/(\tau_S-1) \approx 1.28$, can exist in one model system.

In this paper, we have presented such a model system, which {\color{new} is a  network} of excitatory and inhibitory adaptive Izhikevich neurons~\cite{Izhikevich2003} subjected to stochastic input, 
and shown that a dynamical state of coherent bursting (state II) emerges and displays all the above mentioned features of spontaneous activity. 
We have mapped out the dynamics for the simple random network A of $N=1000$ neurons with homogeneous excitatory and inhibitory incoming degrees 
and constant excitatory and inhibitory synaptic strengths $g_E$ and $g_I$ and found three distinct dynamical states: states I, II and III. 
State II emerges in a range of intermediate values of $g_E$ and its {\color{new} defining} dynamical feature is coherent bursting.
In state II, power-law distributed neuronal avalanches with the observed features,~Eqs.~(\ref{duration})-(\ref{relation}) and $(\tau_T-1)/(\tau_S-1) \approx 1.3$, are found.
These results are also found in the two additional networks studied, network B  with the same network structure as network A but with a uniform distribution of excitatory and inhibitory synaptic strengths centered 
at $g_E$ and $g_I$ and network  C of $N=4095$ neurons with heterogenous excitatory and inhibitory incoming degrees, each ranging from a few to 100, and long-tailed outgoing degree distribution. 

{\color{revised} States I and III are asynchronous irregular states with an approximately time-independent population firing rate,
with state I occurring at small values of $g_E$ and state III at large values of $g_E$, similar to the two types of asynchronous activity found in networks of leaky-integrate-and-fire neurons~\cite{Ostojic2014}. 
As explained in Sec.~\ref{avalanches}, the dynamics of the whole network in states I and III can be approximated by  homogeneous Poisson processes. In both states I and III, the duration distribution of the avalanche, $P(T)$, is exponential while the size distribution, $P(S)$, is approximately exponential for large $S$ and these results are well understood using analytical results derived for a homogeneous Poisson process~\cite{PS2018}.}
{\color{new} The difference between the two states is that distributions of inter-spike interval of individual neurons are well described by exponential distributions  in state I but not in state III, suggesting that the spiking dynamics of individual neurons could further be approximated by homogeneous Poisson processes in state I but not in state III.}
 
Power-law distributed avalanches were reported in stochastic dynamics of networks of Izhikevich neurons with and without synaptic plasticity and were associated with behavior near a critical transition point~\cite{Pasquale2008,KM2019}.
Our study, however, shows that power-law distributed neuronal avalanches are found in the whole regime of state II, and not confined to regions near the transitions between two states. {\color{new} This} suggests that they are likely
caused by the dynamics of state II rather than criticality of the system.
{\color{new} Coherent bursting, the defining} feature of state II, gives rise to an oscillatory population firing rate $R(t)$.
We have shown that the distributions and statistics of the avalanches generated in an inhomogeneous Poisson process of a time varying rate $R(t)$ are in excellent agreement with those found in the network simulation.
This result  {\color{new} thus demonstrates} that the observed power-law distributed neuronal avalanches in state II are consequences of stochasticity and coherent bursting. 
{\color{new} As the avalanches in our inhomogeneous Poisson processes of time varying rate $R(t)$ obey Eq.~(\ref{relation}), they remain to be understood as existing analytical results derived in the slow rate regime~\cite{PRE2017} do not obey Eq.~(\ref{relation}).}
We have further found an interesting result that the conditional size distribution of avalanches for a given duration, $P(S|T)$, obeys approximately a scaling form Eq.~(\ref{scaling}), and this together with a power-law duration distribution, Eq.~(\ref{duration}), can account for all the other three observed features, Eqs.~(\ref{size})-(\ref{relation}). {\color{new} Hence, the problem of explaining the four observed properties of neuronal avalanches is reduced to that of explaining why neuronal avalanches satisfy Eq.~(\ref{scaling}) and have power-law distributed durations.} Further studies are necessary to understand the essential features of $R(t)$ for generating avalanches that have power-law $P(T)$, $P(S|T)$ that obeys the scaling form~Eq.~(\ref{scaling}), and $(\tau_T-1)/(\tau_S-1) \approx 1.3$.

{\color{new} We have approximated the stochastic collective dynamics of the effective neuron in state II by an inhomogeneous Poisson process of the oscillatory firing rate $R(t)$ but this does not imply that the dynamics of individual neurons in state II can be approximated as
$N$ independent inhomogeneous Poisson processes with a common rate $R(t)/N$.} The latter result would imply the former but not vice versa. In particular, the oscillatory $R(t)$ is a result of the interactions among the neurons. With the choice of parameters in the model, 
isolated Izhikevich neurons do not exhibit bursting behavior, thus the observed coherent bursting in state II must be a result of collective dynamics of the whole network due to
interactions among the neurons.
We have shown that in the absence of adaptation, coherent bursting cannot occur and state II does not exist. 
This required balance between excitation and adaptation can be understood as a co-operation between 
 the positive and the negative feedbacks. With such a cooperation, a collective state of coherent bursting can emerge~\cite{VH2001}. 
While the positive feedback is often provided by the excitatory synaptic strength, it is plausible that regulatory mechanisms other than spike frequency adaptation can provide the negative feedback.

In conclusion, we have shown that with a sufficiently strong excitation that is balanced by adaptation, 
coherent bursting that gives rise to oscillations occurs in networks of adaptive neurons, and in the presence of stochastic driving, 
the oscillatory population firing rate resulting from coherent bursting generates power-law distributed neuronal avalanches satisfying Eq.~(\ref{relation}) and $(\tau_T-1)/(\tau_S-1) \approx 1.3$ as observed in experiments. Hence, the observed dynamical features of neuronal avalanches can arise from collective stochastic dynamics of adaptive neurons under 
suitable conditions and need not be signatures of criticality.}
 
\acknowledgements

We thank the anonymous referees for their thoughtful comments which help us to improve the clarity and sharpen the focus of the paper. 

\appendix*
\section{Results for Networks B and C}

\begin{figure}[htbp]
\centering
\includegraphics[width=3.2in]{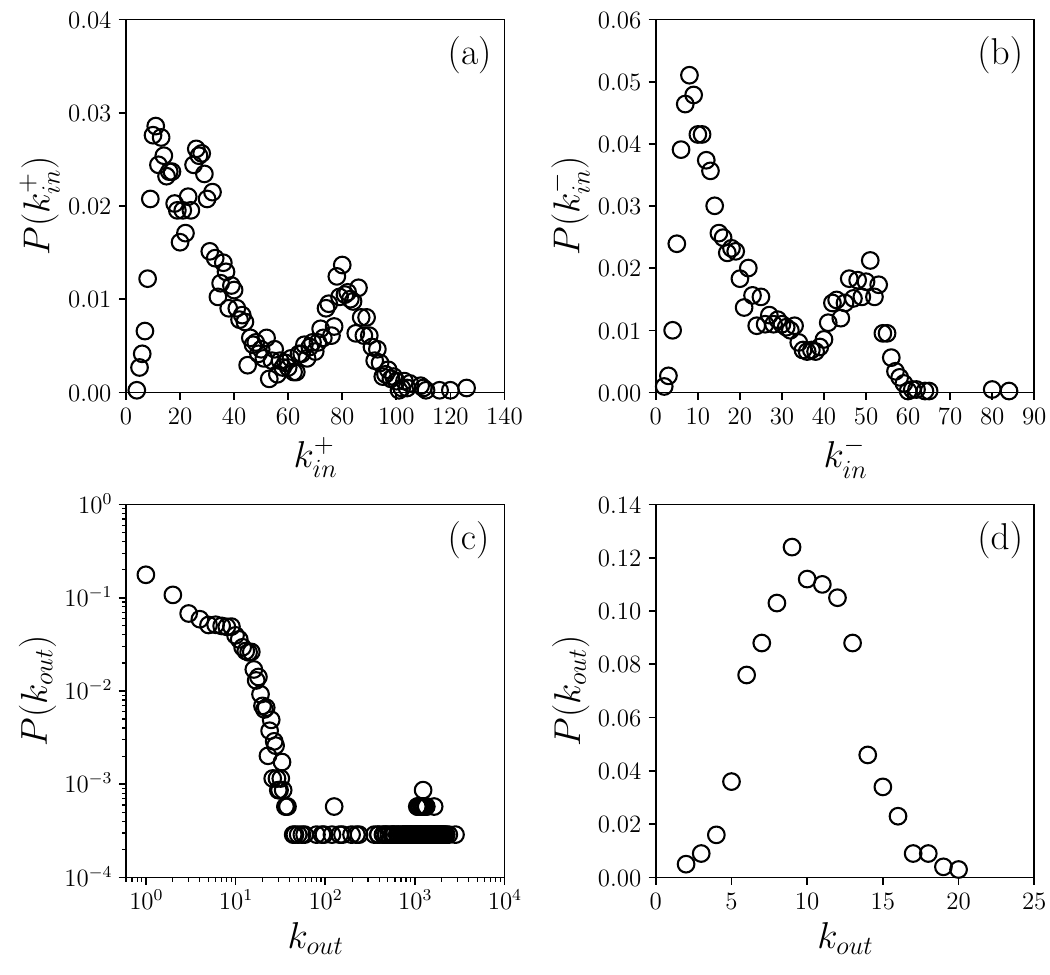}
\caption{Distributions of incoming and outgoing degrees of network C. (a): Distribution of the excitatory incoming degree $k_{in}^+$, (b) Distribution of the  inhibitory incoming degree $k_{in}^-$ and (c) Distribution of the outgoing degree $k_{out}$. 
For comparison, we show the distribution of $k_{out}$ of network A in (d).}
            \label{figA1}
\end{figure}

Network B is modified from network A with the excitatory and inhibitory synaptic strength taken from uniform distributions with mean $g_E$ and $g_I$ and a width of $0.08$.
Network C is reconstructed from multi-electrode array data measured
from a neuronal culture~\cite{RevealPRE2022} using the method of directed network reconstruction proposed by Ching and Tam~\cite{ChingTamPRE2017} and with the excitatory and inhibitory synaptic strengths taken to be constant values $g_E$ and $g_I$. 
Network C has 4095 neurons, more than 4 times that of network A, and the distributions of excitatory and inhibitory incoming degrees are bimodal while the distributions of outgoing degree are long-tailed, which are qualitatively different from those of network A, as shown in Fig.~\ref{figA1}. In particular, in contrast to the homogeneous small incoming degrees of network A, the incoming excitatory and inhibitory degrees are heterogenous and range from a few to around 100 in network C.

\begin{figure}[htbp]
\centering
\includegraphics[width=2.8in]{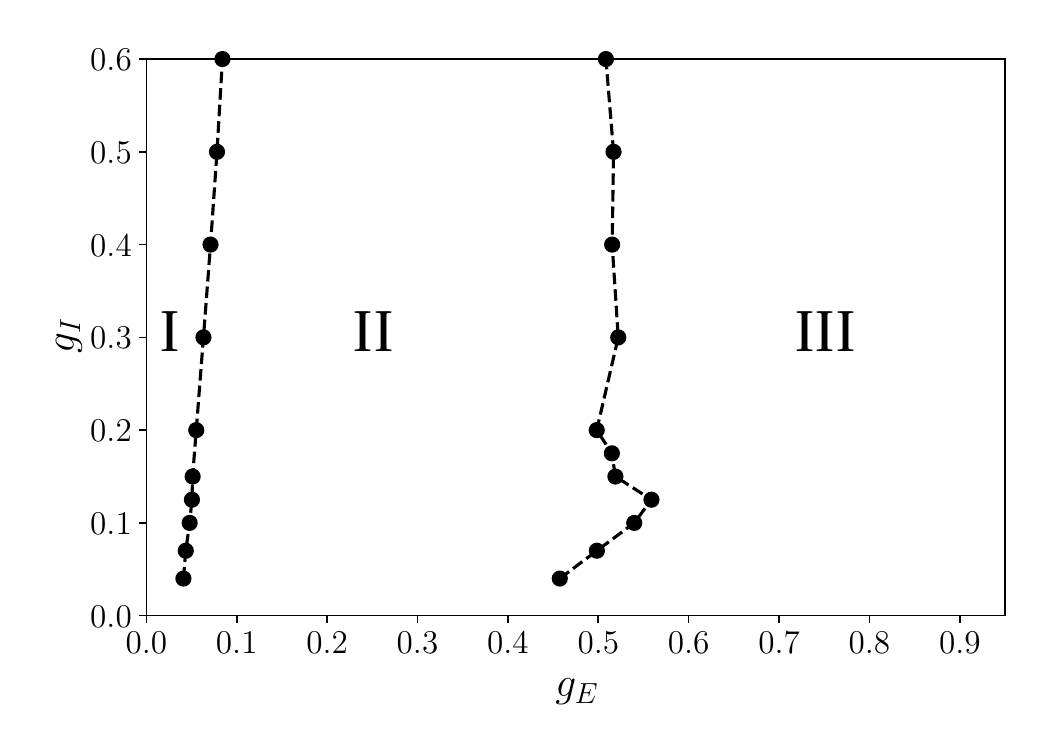}
\caption{Different dynamical states in the $g_E$-$g_I$ parameter space at $\alpha=5$ for network B for $g_E$ and $g_I$ larger than $0.04$, half the width of the uniform distribution.}
                \label{figA2}
\end{figure}

\begin{figure}[htbp]
\centering
\includegraphics[width=3.5in]{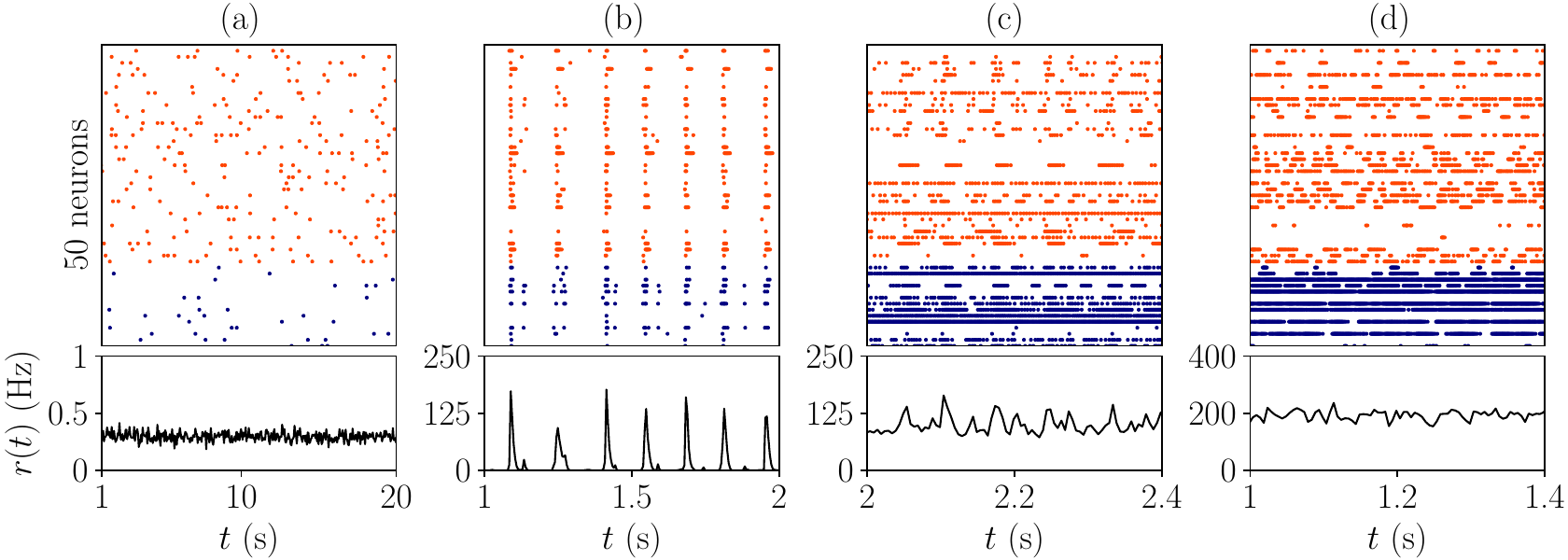}
\caption{Raster plots for 14 inhibitory (blue) and 36 excitatory (red) neurons randomly chosen and the population averaged time-binned firing rate $r(t)$ for $(g_E,g_I)$ equals to
(a)  (0.001875,0.06) in state I, (b) (0.06,0.06) in state II, (c) (0.3, 0.06) in state III 
and  (d) (1.5, 0.06) in state III  for network C. The width of the time bin used in calculating $r(t)$ is 50~ms in (a) and 5~ms in (b), (c) and (d).}
\label{nfigA3} 
\end{figure}

{\color{new} As with} network A, different dynamical states have been found for networks B and C.
The phase diagram obtained for network B is similar to that of network A~(see Fig.~\ref{figA2}). For network C, we show the raster plots as $g_E$ increases at a fixed value of $g_I$ in Fig.~\ref{nfigA3}. States I and II can 
 be clearly seen but because of the heterogeneity of  network C, the participation fraction of neurons in coherent bursting is less than that in network A.
 Moreover, heterogeneity causes larger variations in the raster plot patterns of state III and $r(t)$ exhibits fluctuations in time~(see Fig.~\ref{nfigA3}). 
 For network C, $P(T)$ in state III deviates from an exponential and is not a power-law distribution either.

\begin{figure}[htbp]
\centering
\includegraphics[width=2.8in]{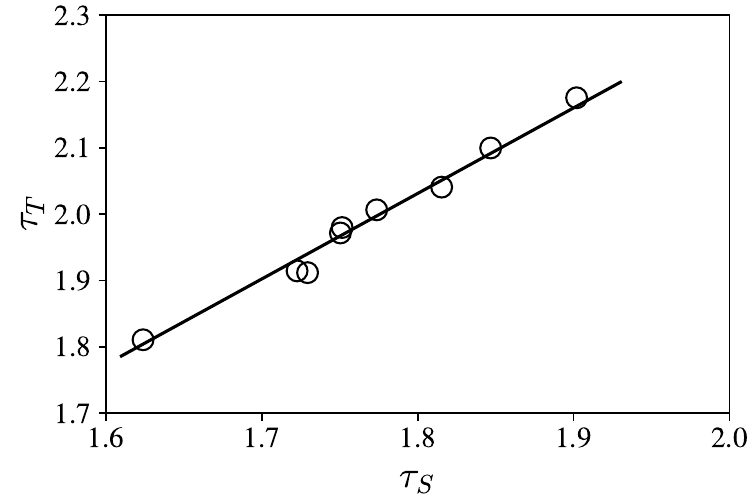}
\caption{Relation between $\tau_T$ and $\tau_S$ in state II of network B for all the cases studied as shown in Table~\ref{ntableA1}. 
{\color{new} The solid line is the least-squares fit $(\tau_T-1)/(\tau_S-1)=K$ and the least-squares estimate of the constant $K$ is $K=1.3$.}}
            \label{nfigA4}
\end{figure}

\begin{figure}[htbp]
\centering
\includegraphics[width=3.0in]{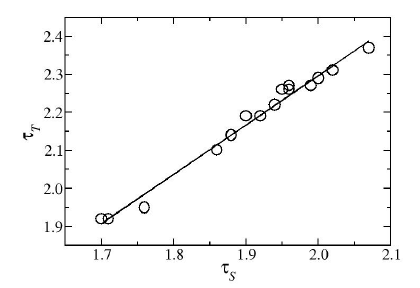}
\caption{Relation between $\tau_T$ and $\tau_S$ in state II of network C for all the cases studied as shown in Table~\ref{tableA1}. {\color{new} The solid line is the least-squares fit $(\tau_T-1)/(\tau_S-1)=K$ and the least-squares estimate of the constant $K$ is $K=1.3$.}}
            \label{nfigA5}
\end{figure}

We focus on the distributions of the neuronal avalanches in state II of networks B and C.
As in network A, the distributions $P(T)$ and $P(S)$ are well described by power law for a range of values of $\Delta t$ and we take $\Delta t={\rm IEI}_{\rm ave}$ or a value around the middle of the range when power-law is not observed at $\Delta t={\rm IEI}_{\rm ave}$. We present the results for the exponents for all the cases studied for networks B and C in Tables~\ref{ntableA1} and~\ref{tableA1}, respectively.
The exponents again obey the scaling relation Eq.~(\ref{relation}) approximately with $(\tau_T-1)/(\tau_S-1)\approx 1.3$ for both networks B and C, as shown in Figs.~\ref{nfigA4} and~\ref{nfigA5}.

\begin{table}[htbp]
    \centering
    \begin{tabular}{|c|c|c|c|c|c|l|} \hline 
            $g_E$ & $g_I$ & $\tau_T$ & $\tau_S$ & $\frac{\tau_T - 1}{\tau_S - 1}$ &  $\gamma$ & $\Delta t \text{ (ms)}$ \\\hline
            0.2&  0.2& 2.18& 1.90& 1.31& 1.31&0.024\\\hline
            0.3& 0.2& 2.04& 1.82& 1.27& 1.29&0.026 (0.017)\\\hline
            0.4& 0.2& 1.91& 1.73& 1.25& 1.29&0.02 (0.013)\\\hline
            0.2& 0.6& 2.01& 1.77& 1.31& 1.27&0.034 (0.025)\\\hline
            0.3& 0.6& 1.98& 1.75& 1.31& 1.31&0.025 (0.014)\\\hline
            0.4& 0.6& 1.81& 1.62& 1.31& 1.27&0.035 (0.009)\\\hline
            0.2& 8& 2.10& 1.85& 1.29& 1.29&0.09 (0.04)\\\hline
            0.3& 8& 1.97& 1.75& 1.29& 1.28&0.04 (0.018)\\\hline
            0.4& 8& 1.91& 1.72& 1.26& 1.26&0.04 (0.012)\\\hline
        \end{tabular}
\caption{Exponents $\tau_T$, $\tau_S$ and $\gamma$ found for different $(g_E,g_I)$ in state II of network B. When $\Delta t \ne {\rm IEI}_{\rm ave}$, the values of ${\rm IEI}_{\rm ave}$ are indicated in parentheses.}
\label{ntableA1}
\end{table}

\begin{table}[htbp]
\begin{tabular}{|c|c|c|c|c|c|c|} \hline 

$g_E$ &  $g_I$ &  $\tau_T$ &  $\tau_S$ &  $\displaystyle \frac{(\tau_T-1)}{(\tau_S-1)}$ & $\gamma$ & $\Delta t$ (ms)\\ \hline 
         0.03&  0.06&  2.37&  2.07& 1.28 & 1.29& 0.02 (0.11)\\ \hline 
         0.0375&  0.06&  2.37&  2.07& 1.28 &  1.28& 0.015 (0.075)\\ \hline 
         0.045&  0.06&  2.26&  1.95&  1.33 &1.29& 0.055\\ \hline 
         0.0525&  0.06&  2.14&  1.88&  1.30 &1.30& 0.04\\ \hline 
         0.06&  0.06&  2.22&  1.94& 1.30 &1.32& 0.035\\ \hline 
         0.0675&  0.06&  2.19&  1.90& 1.32& 1.27& 0.03\\ \hline 
         0.075&  0.06&  2.31&  2.02& 1.28 & 1.25& 0.01 (0.03)\\ \hline 
         0.0975&  0.06&  1.95&  1.76& 1.25 &1.31& 0.025 (0.02)\\ \hline
         0.12& 0.06& 1.92& 1.71&1.30 &1.29&0.02\\ \hline
         0.135& 0.06& 1.92& 1.70& 1.31 &1.27&0.015\\\hline
 0.06& 0.015& 2.29& 2.00& 1.29&1.29&0.04\\\hline
 0.06& 0.03& 2.27& 1.99&1.28 &1.28&0.04\\\hline
 0.06& 0.045& 2.26& 1.96& 1.31  &1.29&0.005 (0.035)\\\hline
 0.06& 0.075& 2.10& 1.86& 1.28& 1.26&0.02 (0.035)\\\hline
 0.06& 0.09& 2.19& 1.92& 1.29 & 1.28&0.025 (0.035)\\\hline
 0.06& 0.105& 2.22& 1.94& 1.30 & 1.26&0.02 (0.04)\\\hline
 0.06& 0.12& 2.27& 1.96& 1.32 &1.27&0.03 (0.04)\\\hline
 \end{tabular}
\caption{Exponents $\tau_T$, $\tau_S$ and $\gamma$ found for different $(g_E,g_I)$ in state II of network C. When $\Delta t \ne {\rm IEI}_{\rm ave}$, the values of ${\rm IEI}_{\rm ave}$ are indicated in parentheses.}
\label{tableA1}
\end{table} 

\end{document}